\documentclass[%
reprint,
superscriptaddress,
amssymb,
aps,
prapp,
longbibliography
]
{revtex4-1}

\usepackage[english]{babel} 
\usepackage [table]{xcolor}

\usepackage{graphicx, upgreek} 

\usepackage{amsmath}
\usepackage{amsthm}
\usepackage{amsfonts}
\usepackage{color}
\usepackage{natbib}

\usepackage{float}
\usepackage{textcomp}
\usepackage{graphicx}
\usepackage{dcolumn}
\usepackage{bm}
\usepackage[mathlines]{lineno}
\usepackage[normalem]{ulem} 

\usepackage{array}

\usepackage[colorinlistoftodos]{todonotes}

\begin{document}


\title{Tunable Nb superconducting resonators based upon a Ne-FIB-fabricated constriction nanoSQUID}
\author{O.~W.~Kennedy}
\affiliation{London Centre for Nanotechnology, UCL, 17-19 Gordon Street, London, WC1H 0AH, UK}
\author{J.~Burnett}
\affiliation{London Centre for Nanotechnology, UCL, 17-19 Gordon Street, London, WC1H 0AH, UK}
\affiliation{Microtechnology and Nanoscience, Chalmers University of Technology, SE-41296 Gothenburg, Sweden.}
\author{J.~C.~Fenton}
\affiliation{London Centre for Nanotechnology, UCL, 17-19 Gordon Street, London, WC1H 0AH, UK}
\author{N.~G.~N.~Constantino}
\affiliation{London Centre for Nanotechnology, UCL, 17-19 Gordon Street, London, WC1H 0AH, UK}
\author{P.~A.~Warburton}
\affiliation{London Centre for Nanotechnology, UCL, 17-19 Gordon Street, London, WC1H 0AH, UK}
\affiliation{Dept.\ of Electronic and Electrical Engineering, UCL, London, WC1E 7JE, UK}
\author{J.~J.~L.~Morton}
\affiliation{London Centre for Nanotechnology, UCL, 17-19 Gordon Street, London, WC1H 0AH, UK}
\affiliation{Dept.\ of Electronic and Electrical Engineering, UCL, London, WC1E 7JE, UK}
\author{E.~Dupont-Ferrier}
\affiliation{London Centre for Nanotechnology, UCL, 17-19 Gordon Street, London, WC1H 0AH, UK}
\affiliation{Department de Physique, Institut Quantique, Universit\'{e} de Sherbrooke, 2500 Boulevard de l`Universit\'{e}, Sherbrooke, Quebec J1K 2R1, Canada}

\begin{abstract}
Hybrid superconducting--spin systems offer the potential to combine highly coherent atomic quantum systems with the scalability of superconducting circuits. To fully exploit this potential requires a high quality-factor microwave resonator, tunable in frequency and
able to operate at magnetic fields optimal for the spin system.
Such magnetic fields typically rule out conventional Al-based Josephson junction devices that have previously been used for tunable high-$Q$ microwave resonators. 
The larger critical field of niobium (Nb) allows microwave resonators with large field resilience to be fabricated.
Here, we demonstrate how constriction-type weak links, patterned in parallel into the central conductor of a Nb coplanar resonator using a neon focused ion beam (FIB), can be used to implement a frequency-tunable resonator. 
We study transmission through two such devices and show how they realise high quality factor, tunable, field resilient devices which hold promise for future applications coupling to spin systems.

\end{abstract}

\selectlanguage{english}

\date{\today}

\maketitle

\section{Introduction}

A large and growing variety of spin systems have been coupled to superconducting resonators, including ensembles of non-interacting spins in silicon~\cite{bienfait2016controlling}, diamond~\cite{kubo2010strong} and other materials~\cite{probst2013anisotropic}, magnons in ferrimagnets~\cite{huebl2013high} and chiral magnetic insulators~\cite{abdurakhimov2018strong}, and individual spins in quantum dot devices~\cite{mi2016strong}. Motivations for such studies include long-range coupling of spin qubits~\cite{grezes2016towards}, realisation and study of topological systems~\cite{Larsen2015}, long-lived microwave quantum memories for superconducting qubits~\cite{yan2016flux,Dunsworthloss}, and the demonstration of microwave-to-optical conversion at the single-photon level~\cite{Blum2015}. Planar superconducting circuits provide a robust, well-studied~\cite{burnett2014evidence} and scalable architecture~\cite{barends2014superconducting} for such hybrid systems, and superconducting resonators with $Q$-factors over 1 million have been achieved~\cite{megrant2012planar}. However, for the majority of the applications described above, externally applied magnetic fields from $\sim$10~mT up to several 100~mT, or more, are required to bring the spin systems into a suitable regime of interest. Furthermore, control of the spin-resonator coupling is required, often on short timescales, in applications such as quantum memories, and may be achieved, for example, by frequency-tuning of the resonator.


Superconducting quantum interference devices (SQUIDs) act as flux-tunable inductors and have been successfully incorporated into resonators to provide frequency tunability~\cite{palacios2008tunable,sandberg2008tuning,levenson2011nonlinear}. The SQUID inductance is tuned from its minimum to its maximum by a change of half a flux quantum in the flux threading the SQUID loop, thus altering the resonator frequency. This means that small local fields provided by on-chip flux lines are able to tune SQUID-embedded resonators on timescales of a few nanoseconds~\cite{sandberg2008tuning}. Technologies which use DC currents to tune the kinetic inductance and hence resonant frequency of resonators have also been developed~\cite{adamyan2016tunable, asfaw2017multi}and coupled to spin ensembles \cite{kubo2010strong}. Previous SQUID-tunable devices have been fabricated from aluminium with shadow-evaporated junctions~\cite{sandberg2008tuning} or used Nb/Al/AlO$_x$/Nb trilayer junctions~\cite{palacios2008tunable}. Al devices may suffer from the low critical field of Al and are expected to have poor magnetic-field resilience. AlO$_x$ tunnel junctions may introduce extra losses to resonators and limit quality factors. An alternative technology is based on nanoSQUIDs \cite{granata2016nano, martineznanosquids} formed by superconducting-nanowire-based constriction junctions; these have already been shown to possess exceptional field resilience~\cite{schwarz2013}. 

NanoSQUIDs are commonly fabricated~\cite{hao2009characteristics} by a Ga-based focused ion beam (FIB); however, this technique has been shown to induce loss into superconducting resonators~\cite{jenkins2014nanoscale}. Here, we use a neon FIB (a technique shown to be compatible with high quality superconducting resonators~\cite{burnettnanowireloss}) to create constrictions in the central conductor of a Nb superconducting $\lambda/4$ co-planar resonator. These constrictions have a width of 50~nm and are placed in parallel such that they complete a superconducting loop between the current antinode of the resonator and the ground (Fig.~\ref{fig:device_pic}). We study the microwave transmission of two such devices fabricated from Nb films of different quality as determined by measuring the quality of bare resonators fabricated from the same films. One such device includes a lumped element inductor which reduces the tunability of the resonator but may also be used to couple more strongly to spins in future experiments. We demonstrate that nanoSQUID-embedded resonators may realize high-quality, frequency-tunable and field-resilient resonators. 

\section{Experimental}
Co-planar resonators were fabricated by etching thin films of superconductor using a similar method to that described in Ref~\cite{burnettnanowireloss}. Here Nb is used, instead of NbN, because of its longer coherence length of 38~nm~\cite{meservey1969equilibrium} which sets the lengthscale for the width of constrictions needed to make junctions --- Nb thus allows $\approx$50~nm constrictions to be used, which is easier to achieve than the $\approx$20~nm constrictions required in NbN. Nb also has a lower kinetic inductance (hence lower impedance and larger zero-point current fluctuations), which could enable stronger magnetic coupling of the Nb resonator to spins. Nb films were deposited on Si substrates by DC magnetron sputtering from a 99.99\%-pure elemental Nb target in argon. The pressure before deposition was 6$\times 10^{-7}$~mbar and during deposition was 3.5$\times 10^{-3}$~mbar. The sputter power was 200~W, with the deposition timed to produce a 50-nm-thick film. 

\begin{figure}
	\centering
		\includegraphics[width=0.45\textwidth]{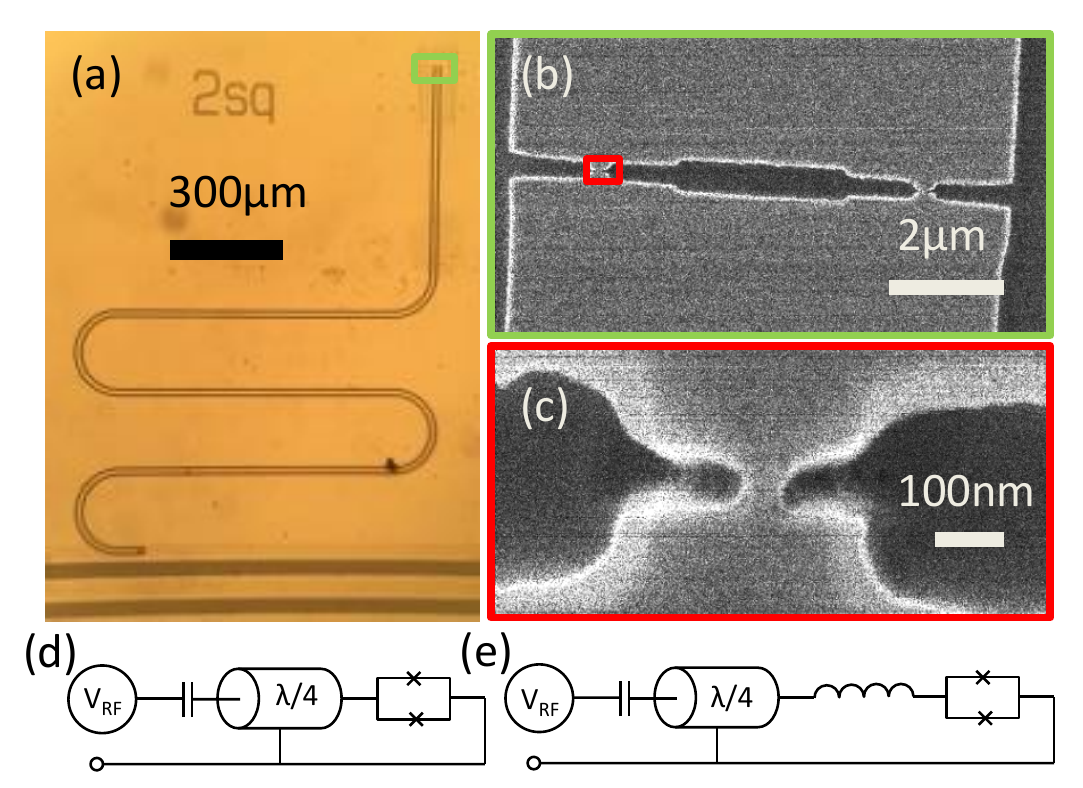}
	\caption{(a) Optical micrograph of resonator ATune (see text) with the microwave feedline shown at the bottom. The green rectangle indicates the location of (b), a He FIB micrograph, collected using an electron flood gun, showing the SQUID loop at the grounded end of ATune. The red rectangle indicates the location of (c) Ne-FIB-milled constriction with a width of 50~nm. In (a), (b) and (c), the lighter tones show the Nb and the darker tones the Si. (d, e) Equivalent circuit diagrams of the tunable resonators for (d) ATune and (e) BTune. The lumped element inductor is present in BTune and not in ATune.}
	\label{fig:device_pic}
\end{figure}

Two chips are fabricated from Nb films deposited according to the same recipe in separate sputter runs and labelled A and B. We discuss two resonators from each of these chips, one tunable and one bare resonator labelled Tune and Bare respectively. We refer to resonators by a combination of chip~And resonator label e.g. ATune is the tunable resonator from chip~A. chip~A (B) has an area of 58~mm$^2$ (44~mm$^2$). Before deposition, chip~B is dipped into HF for $\sim$10~s to remove a native oxide coating and then loaded into the sputter system and put under vacuum within 5-10~minutes. Quarter-wave ($\lambda$/4) resonators (see Fig.~\ref{fig:device_pic}a) with an embedded superconducting loop at the grounded end of the resonator were patterned by electron beam lithography (EBL) (Fig.~\ref{fig:device_pic}b) in the same lithography step as a microwave feed line. The resist pattern was transferred into the film by a reactive ion etch (RIE) process using a 2:1 ratio of SF$_6$ to Ar, at 30~mbar and 30~W for 120~s. The RIE process additionally etches exposed Si to a depth of 500~nm, leaving areas with Nb raised above their surroundings. Resonators on chip~B have a long constriction (100~$\upmu$m $\times$ 1 $\upmu$m) embedded into the central conductor just before the SQUID. The narrow constriction create larger local fluctuating fields and increase the resonator coupling to spins in future devices \cite{bienfait2016controlling}. The increased coupling comes at the cost of including a constriction which acts as a lumped element inductor reducing the inductance ratio between SQUID and resonator, which consequently reduces the resonator tunability.

The superconducting loop has two constrictions (see Fig.~\ref{fig:device_pic}b and c): broad constrictions are defined in the initial EBL exposure and subsequently narrowed to $\approx$ 50~nm by Ne FIB milling, in which a beam of Ne ions, accelerated to 15~kV, mills through the Nb. A dose of $\approx$2~nC$\upmu$m$^{-2}$ is used. Circuit diagrams of these devices are shown in Fig.~\ref{fig:device_pic} d (e) for Atune (BTune). On chip~A, 21 out of 22 constrictions milled were still intact after narrowing to a dimension approaching the coherence length, suggesting a high yield for this part of the processing; see supplementary materials \cite{SI} for statistical analysis on junction widths.

Resonators are measured at a temperature $T\approx300$~mK in a $^3$He cryostat with a heavily attenuated microwave in-line and a cryogenic high-electron-mobility transistor (HEMT) amplifier on the microwave out-line. The $S_{21}$ transmission of microwaves through the feed-line ---to which the resonator is capacitively coupled--- is measured using a Rohde \& Schwarz ZNB8 vector network analyzer. Perpendicular magnetic fields are applied by a superconducting magnet connected to a precision current source (Keithley 2400 SourceMeter). Samples are enclosed in a brass box lined with Eccosorb CR-117, a microwave absorber, to reduce the number of quasiparticles excited by stray IR photons, which we have previously shown can have a significant effect on superconducting constrictions~\cite{burnettnanowireloss}.

\section {Results}
\begin{figure}
	\centering
	\includegraphics[width=0.5\textwidth]{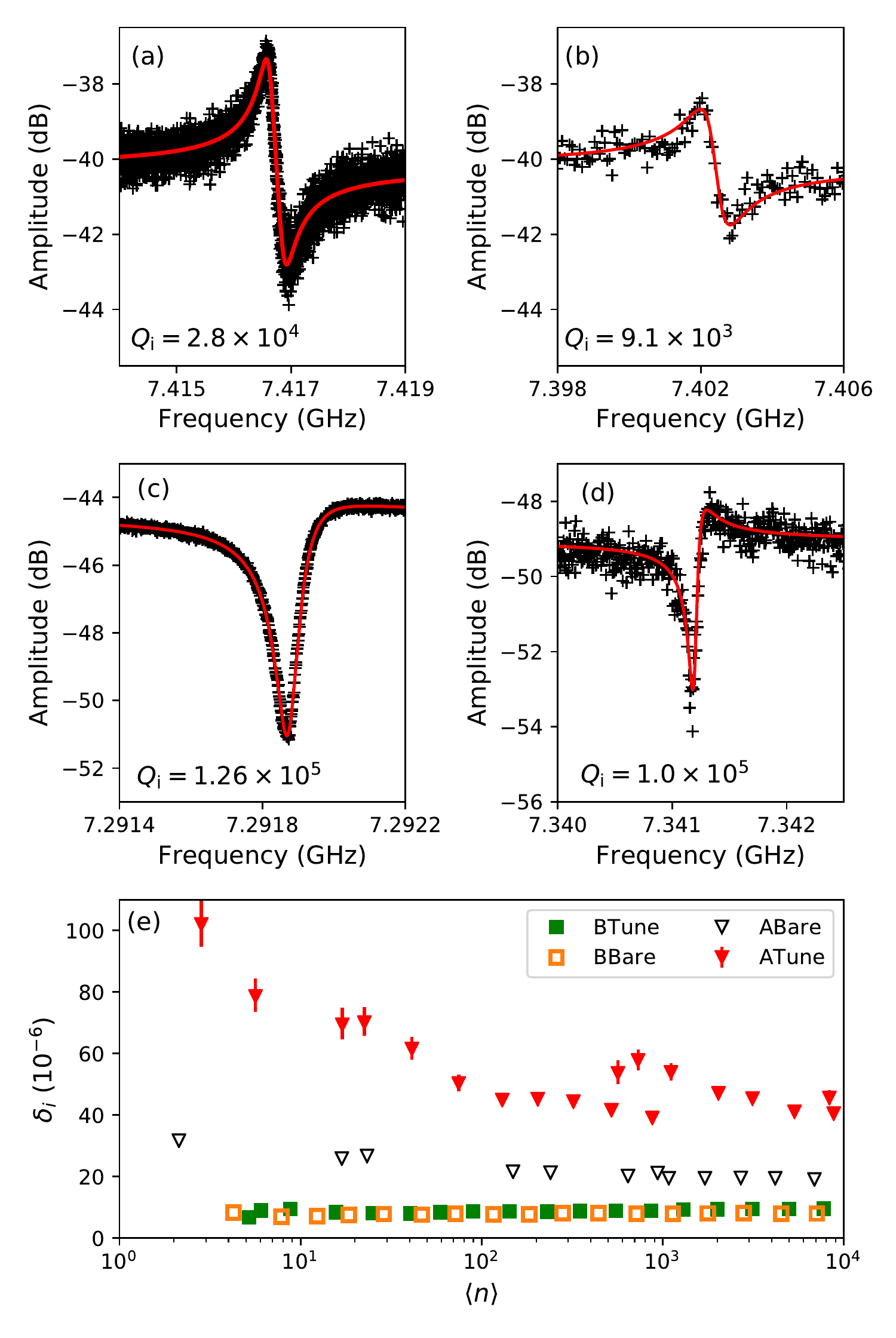}
	\caption{S$_{21}$ amplitude response (black crosses) and fits (red line) for (a, b) ATune and (c, d) BTune. Measurements are performed at (a, c) zero field and (b, d) $\sim 0.5$~mT applied perpendicular field. All responses are recorded at a power of $\sim -106$~dBm and internal quality factors are indicated in the figure. The lower frequency of BTune in (c) than in (d) is due to device aging between measurements. The local fields in (b) and (d) are $\sim$60~mT and $\sim$20~mT respectively. (e) Internal losses of all four resonators as a function of average photon number stored in the resonator. Error bars are shown where they are larger than the markers. }
	\label{fig:resfit}
\end{figure}

We first characterize the zero-field characteristics of the resonators on the two chips at the base temperature of the cryostat. In Fig.~\ref{fig:resfit}, we show the zero-field magnitude response of $S_{21}$ from ATune (Fig.~\ref{fig:resfit}~a) and BTune (Fig.~\ref{fig:resfit}~c) at an applied power of $\sim$-106~dBm. Using a traceable fit routine~\cite{doi:10.1063/1.4907935}, based on fitting a circle to the resonance in the real--imaginary plane, we extract the resonator parameters (fits shown in red in Fig.~\ref{fig:resfit}). At an applied power of $\sim-106$~dBm ATune, ABare, BTune and BBare have internal quality factors $Q_{\rm i} = 2.8 \times10^4$, $5.3\times10^4$, $1.25\times10^5$ and $1.26\times10^5$ respectively (resonance notches of bare resonators are not shown). The biggest difference between zero-field quality factors is between the different films rather than between the bare and tunable resonators. This shows that at high drive power, the SQUIDs introduce minimal extra losses and that the losses are dominated by effects arising from the film quality. The asymmetry of the resonance of ATune (Fig.~\ref{fig:resfit}a, b) persists down to single-photon powers due to an impedance mismatch and this is fully captured by the fit routine. Fig.~\ref{fig:resfit}b, d shows that both ATune and BTune resonances are present at applied perpendicular fields approaching 0.5mT with resonance responses. We discuss resonator behaviour in an applied magnetic field later in the manuscript.

We measure the internal losses of all four resonators ($\delta_i = 1/Q_i$) as a function of the average photon number within the resonator (see Fig.~\ref{fig:resfit}~e). Resonators from film A show higher losses across all powers than resonators from film B irrespective of whether they are bare or tunable. Whilst ATune shows more losses than ABare and a greater increase in losses at low power, for film B the losses are approximately equal for BTune and BBare. We therefore attribute the difference in losses in film A to on-chip variation in film quality, as commonly seen in CPW resonators. Losses are dominated by losses associated with the film rather than losses introduced by the SQUID even in the single photon limit. The difference between films could be a result of the small recipe differences such as the HF dip for film B; however, the increased loss at high powers is indicative of increased conductor losses associated with the quality of superconducting films.  

\subsection{Tunability}

\begin{figure*}
    \centering
    \includegraphics[width=\textwidth]{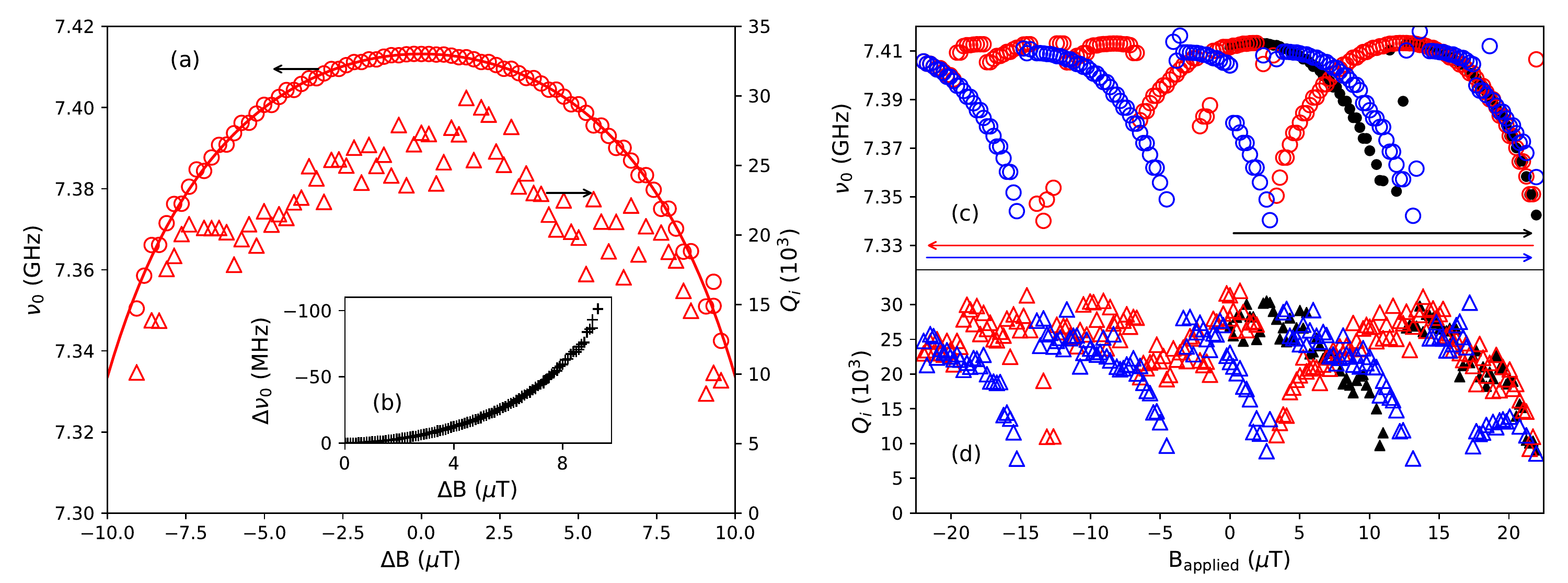}
    \caption{
    (a) The tuning of resonant frequency (circles) and quality factor (triangles) of ATune as magnetic field is swept. The resonant frequency is fitted by Eq.~\ref{eq:nu}, with fitting parameters $A=0.050$ and $K=0.101$ ($\upmu$T)$^{-1}$. For the data shown, the central value $\Delta{B}=0$ corresponds to an applied field $B=12.4~\upmu$T.
    (b) Tuning of the resonator when the field is changed in 0.1$\upmu$T steps from $B=12.3$~$\upmu$T. A maximal detuning of the resonator of $-101$~MHz is obtained.
    (c) Resonant frequency and (d) internal quality factor of ATune as applied field is swept from 0 to 22~$\upmu$T (black), then down from 22~$\upmu$T to $-22~\upmu$T (red) and then back up from  $-22~\upmu$T to 22~$\upmu$T (blue). The sweep directions are shown by arrows at the bottom of (c). 
    }
    \label{fig:up_down_single_period}
\end{figure*}

In order to study resonator tuning we examine ATune. Without a lumped element inductor, the tunable SQUID inductance is a greater fraction of the total inductance and so this resonator tunes more in frequency.
We first study its behaviour in a small perpendicular magnetic field ($\leq 10~\upmu$T). Fig.~\ref{fig:up_down_single_period}~a shows a typical field sweep; this reveals the smooth tuning of the resonator towards lower frequencies 
as magnetic field amplitude is increased. The full range of frequency tuning is found to be  $\approx$100~MHz, obtained by changing the applied perpendicular field by $\approx$10~$\upmu$T (see Fig.~\ref{fig:up_down_single_period}~b) \footnote{Here, field has been changed in small (0.1~$\upmu$T) steps to approach maximal detuning whilst ensuring that the point of maximal detuning is not overshot before the sweep direction is reversed.}. 
To analyse this behaviour, we start by assuming a Josephson-like sinusoidal current-phase relationship for the nano-constrictions since the length:width ratio of the constrictions shown in Fig.~\ref{fig:device_pic}~c is approximately one \cite{golubov2004current}. We therefore treat the superconducting loop as a DC-SQUID with a flux-tunable inductance \cite{sandberg2008tuning}
\begin{equation}
L_{\rm SQUID} = \Phi_0/[4\pi I_{C0}|\cos{f}|],
\label{eq:Lsq}
\end{equation}
where $I_{\rm C0}$ is the zero-field critical current of the SQUID, $f = \pi\Phi/\Phi_0$ is the frustration of the SQUID and $\Phi_0$ is the flux quantum.

The total impedance $Z_T$ of a transmission line terminated by a SQUID is given at frequency $\nu$ by 
\begin{equation}
Z_T=\frac{Z(Z_{\rm SQUID}+jZ\tan{(2\pi\nu d/v)})}{(Z+jZ_{\rm SQUID}\tan{(2\pi\nu d/v}))},
\label{eq:ZT}
\end{equation}
where $d$ is the length of the transmission line, $Z$ is the impedance of the transmission line, $Z_{\rm SQUID}$ the impedance of the SQUID and  $v = \sqrt{1/l_0c_0}$ the speed of light in the transmission line where $c_0$ ($l_0$) is the capacitance (inductance) per unit length. $Z_T$ is real at resonance and the resonant frequencies $\nu_i$ are therefore given by
\begin{equation}
\tan{\bigg(\frac{2\pi\nu_i d}{v}\bigg)}=\frac{|Z_T|}{l_0v},
\label{eq:res-term}
\end{equation}
which may be solved numerically. The fundamental resonant frequency may be expressed approximately [\citenum{jones2013tunable}] in terms of the total inductance $L$ and capacitance $C$ of the distributed resonator:
\begin{equation}
\nu_{0}(B) = \frac{1}{4\sqrt{(L_{\rm res} + L_{\rm SQUID}^{0} + \Delta L(B))C}},
\end{equation}
where $L_{\rm res}$ is the inductance of the resonator excluding the SQUID, $L_{\rm SQUID}^{0}$ is the zero-field inductance of the SQUID.  $\Delta L(B)$ is the change of inductance of the SQUID with field which, from Eq.~\ref{eq:Lsq}, is equal to $\Phi_0/4\pi I_{\rm C0}\times(1/|\cos{(f)}| - 1)$. 
Assuming $f \propto B$, an assumption which we examine further below, we can write

\begin{equation}
\frac{\nu_{0}(B)}{\nu_{0}(0)} = \sqrt{\frac{1}{1 + \frac{\Delta L(B)}{L_{res} + L_{\rm SQUID}^{0}}}} = \sqrt{\frac{1}{1 - A + \frac{A}{|\cos{(KB)}|}}},
\label{eq:nu}
\end{equation}
where $A = \Phi_0/[4\pi I_{\rm C0}(L_{\rm res} + L^{0}_{\rm SQUID})]$ and $f=KB$ so that $K$ scales field to $f$. The observed $\nu_0(B)$ dependence of the resonator in Fig.~\ref{fig:up_down_single_period} fits well with Eq.~\ref{eq:nu}, allowing determination of $A$ and $K$.

We next consider the relation between the field periodicity of the tuning behaviour and the flux quantum. SQUID behaviour is periodic in applied flux. The area of the SQUID loop in ATune is $A_{\rm loop}=3.7\pm$0.3$~\upmu$m$^2$, so 10~$\upmu$T (the field required to maximally tune the resonator) corresponds to a flux $BA_{\rm loop}\approx 0.02\Phi_0$. Assuming the tuning arises from the SQUID, the field required to maximally tune the resonator implies that the local flux density at the SQUID is much greater than the 10~$\upmu$T applied by the magnet. This indicates substantial flux focusing due to flux expulsion from the superconducting ground plane surrounding the resonators. We return to the topic of flux focusing later.

As ATune is tuned away from $\nu_{0}$, $Q_{\rm i}$ is found to drop from its maximum value, 2.8$\times$10$^4$, to  1.0$\times$10$^4$ when maximally tuned. The same trend of $Q_{\rm i}$ decreasing with tuning is also seen in BTune. This phenomenon of $Q_{\rm i}$ decreasing with tuning has previously been observed and attributed to increasing thermal noise as the SQUID is tuned~\cite{palacios2008tunable} or increased dissipation caused by a sub-gap resistance~\cite{sandberg2008tuning}. 
An alternative explanation could be that dilute surface spins~\cite{degraaf2017spin} induce spectral broadening of the resonance lineshape by flux-noise-based frequency jitter in these flux-tunable resonators. However, even for the highest values of flux noise in Ref.~\citenum{ithier2005}, which correspond to $\sim$100$\upmu\Phi_{0}$, the corresponding frequency jitter would be too small to create sufficient spectral broadening to explain the drop in $Q_{\rm i}$ observed here. The source of these extra losses is the subject of ongoing work.


\subsection{Hysteresis and Premature Switching}

When tuning ATune over more than one period, as shown in Fig.~\ref{fig:up_down_single_period}c and d, a hysteretic behaviour is seen, similar to that previously reported in Al constriction SQUID resonators \cite{levenson2011nonlinear}. In addition, frequency jumps are observed at values of tuning less than the maximum value (see for example the region between $-5$~$\upmu$T and $-20$~$\upmu$T). We attribute jumps at non-maximal tuning to flux trapping as the field is ramped up and down.
The oscillations in the internal quality factor (Fig.~\ref{fig:up_down_single_period}d) follow the frequency tuning of the resonator, showing that the degradation in $Q_{\rm i}$ with magnetic field arises from the state of the SQUID and not from the properties of the resonator in a magnetic field. 

\begin{figure}
    \centering
    \includegraphics[width=0.5\textwidth]{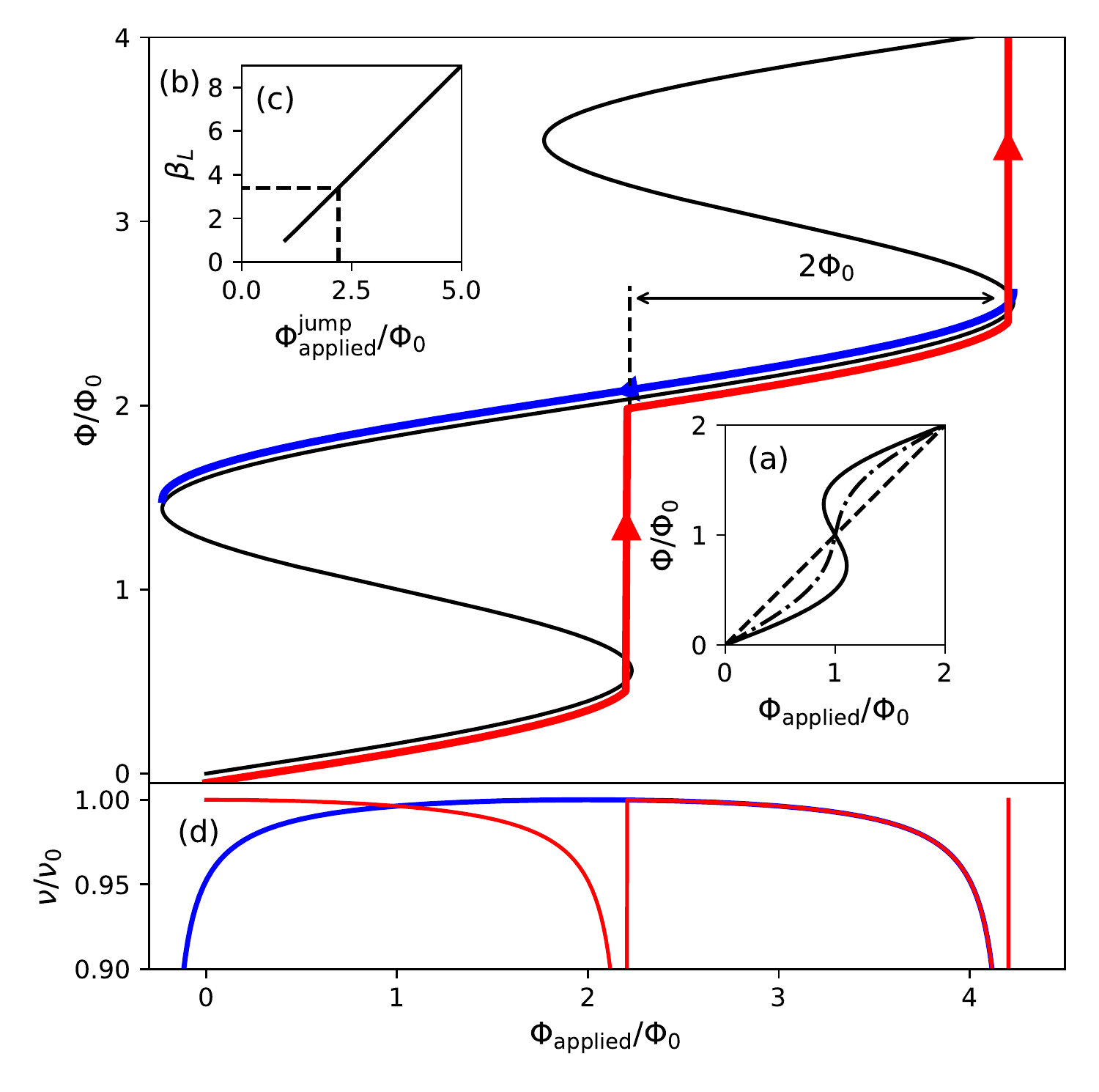}
    \caption{(a) Variation with applied external flux $\Phi_{\rm applied}$ of flux $\Phi$ threading the SQUID, for SQUIDs with $\beta_{\rm L} =$0 (dashed), 0.5 (dot-dashed) and 1 (solid) lines. (b) Variation with applied flux of flux threading a $\beta_{\rm L}$=3.4 SQUID. The black line shows the solution to the equation governing flux threading a hysteric SQUID \cite{tinkham2004introduction}. The red (blue) line indicates how the flux threading the SQUID evolves in the absence of fluctuations with the applied flux ramping upwards (downwards) as in Fig.~\ref{fig:up_down_single_period}c,d. (c) The relationship between applied flux required to maximally detune resonator and $\beta_{\rm L}$ of the embedded SQUID. (d) Variation of resonant frequency of the SQUID with applied flux depending on how field is ramped, calculated using an inductance ratio between SQUID and resonator determined from the parameter $A$ from the fit in Fig.~\ref{fig:up_down_single_period} (a). Colors of lines correspond to the same flux ramping as in (b).}
    \label{fig:beta_L_jumps}
\end{figure}

The hysteretic tuning of the resonant frequency and internal quality factor may be explained by significant self-inductance of the superconducting SQUID loop. 
The SQUID has a characteristic parameter $\beta_{\rm L} =  2L_{\rm loop}I_{\rm C}/\Phi_0$ (where $L_{\rm loop}$ is the inductance of the loop and $I_{\rm C}$ is the Josephson critical current). 
%
%
When $\beta_{\rm L}\gtrsim 1$, the SQUID behaviour becomes hysteretic with applied flux, as shown in Fig.~\ref{fig:beta_L_jumps}a. The red path in Fig.~\ref{fig:beta_L_jumps}b maps out the flux within the SQUID as the field is increased (assuming zero temperature, in the absence of fluctuations). At extremal points, the flux threading the SQUID exhibits discontinuous jumps  as $\Phi_{\rm app}$ is ramped upwards, occurring periodically with a period of 2$\Phi_0$. At finite temperature, thermal fluctuations cause these jumps to occur at a temperature-dependent flux less than that at $T=0$.

\begin{figure*}
    \centering
    \includegraphics[width=\textwidth]{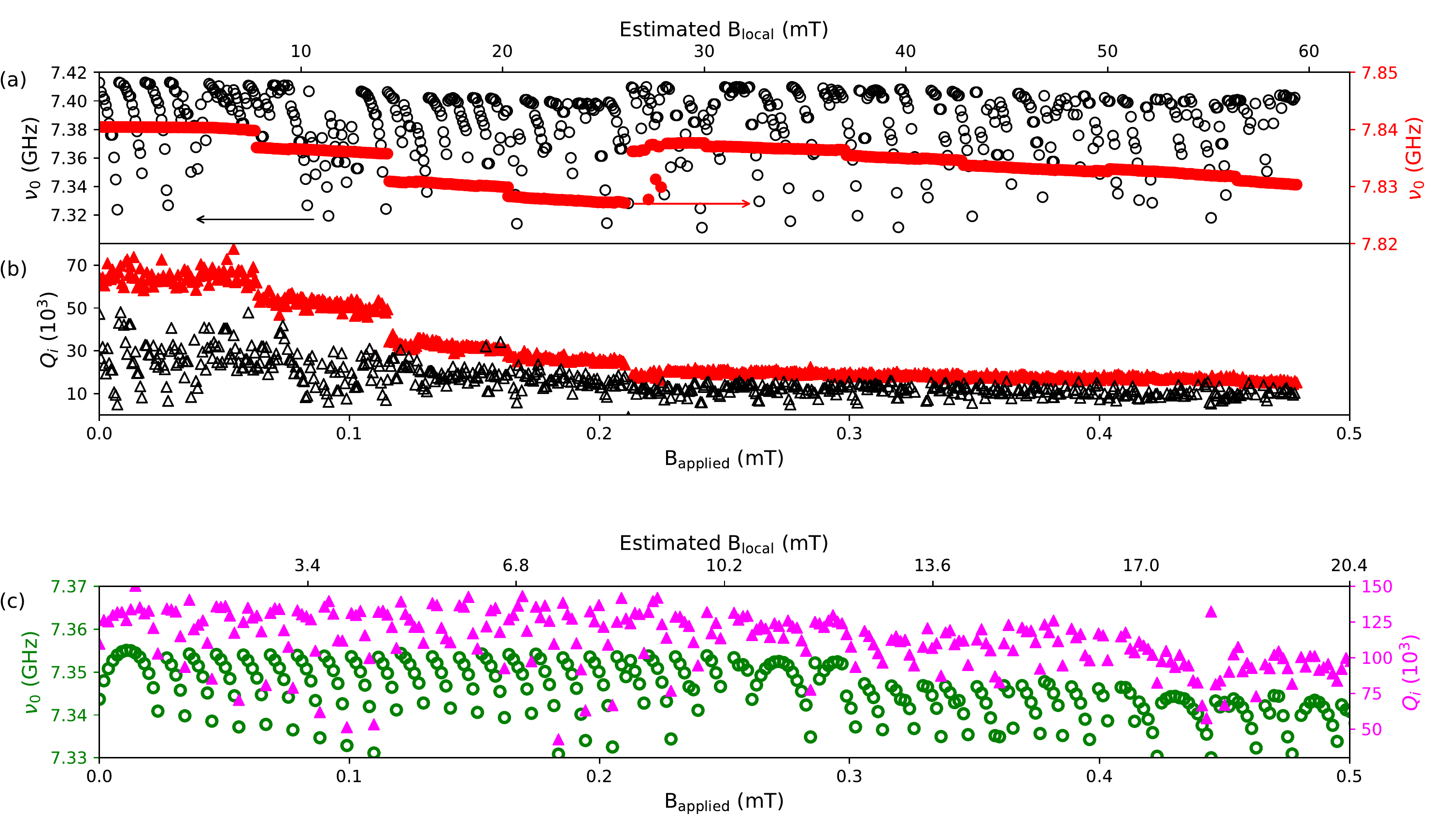}
    \caption{Magnetic-field dependence of (a) resonant frequency and (b) internal quality factor for ATune (black, unfilled symbols) and ABare (red, filled symbols). The local field $B_{\rm local}$ is estimated based on measurements of flux focusing (see main text). (c) Magnetic field dependence of resonant frequency (unfilled green circles) and internal quality factor (magenta triangles) for BTune. Local field is determined as for ATune.}
    \label{fig:large_field}
\end{figure*}

Using this 2$\Phi_0$ periodicity,  we are able to calibrate local fields at our device and experimentally quantify flux focusing in these hysteretic devices. Jumps occur every 9.7~$\upmu$T (averaged over 4 consecutive jumps in Fig.~\ref{fig:large_field}), which corresponds to $BA_{\rm loop}\approx$0.016$\Phi_0$. Identifying the jumps as $2\Phi_0$-periodic features, we infer a flux-focusing $\mathcal{F}\approx 124$, where we have defined $\mathcal{F}$ by $B_{\rm local} = \mathcal{F}B$.
Significant flux focusing from superconducting ground planes has recently been investigated theoretically and experimentally in Ref.~\citenum{bothner2017improving}, where simulations gave $\mathcal{F} \approx 27.5$. The extent of flux focusing is specific to device geometry; for example, features designed to trap flux can have large effects on $\mathcal{F}$.

The change in applied flux required to maximally tune the resonator frequency is determined by the $\beta_{\rm L}$ value of the SQUID (see Fig.~\ref{fig:beta_L_jumps}c).
At finite temperature, the flux in the SQUID jumps before reaching the point of instability shown in Fig.~\ref{fig:beta_L_jumps}, and so the experimentally measured jump positions provide a lower bound on $\beta_{\rm L}$. We thus infer that $\beta_{\rm L}>3.4$ for our device.
Using Eq.~\ref{eq:Lsq} and the fit parameter $A$ (which relates the inductance of the resonator and of the SQUID), we can calculate the expected tuning of the resonator in an applied magnetic field based on a sinusoidal current--phase relation (Fig.~\ref{fig:beta_L_jumps}d). The smooth tuning shown in Fig.~\ref{fig:up_down_single_period}a (blue line in Fig.~\ref{fig:beta_L_jumps}b,d) and jumps seen in Fig.~\ref{fig:up_down_single_period}c (red line in Fig.~\ref{fig:beta_L_jumps}b,d) are qualitatively reproduced. 
The calculation, however, suggests that resonant frequencies should decrease significantly in the vicinity of hysteretic jumps due to the asymptote in $1/\cos{f}$ at $f=1/2$, whereas in practice these resonators tune by only $\sim 1\%$ of their untuned frequency.
Numerical analysis~\cite{kim2015circuit} based on Eq.~\ref{eq:res-term} predicts such reduced tuning for resonators where the SQUID inductance or capacitance is a significant fraction of the total inductance or capacitance of the distributed resonator. We determine the untuned inductance of the SQUID to be approximately 10\% of the total inductance of the distributed resonator (see supplementary materials \cite{SI} for details), which in Ref.~[\citenum{kim2015circuit}] is sufficient (even with small SQUID capacitance) to reduce the tuning of the resonator approaching $\Phi = \Phi_0/2$ to only around 30\%. Additionally, asymmetry between junctions allows switching at smaller tuning than perfectly symmetric devices as the narrower junction becomes maximally biased (and hence switches) before the wider junction becomes maximally biased.

\subsection{Magnetic field resilience}

In Fig.~\ref{fig:large_field}, ATune (black markers), ABare (red markers) and BTune (green and magenta markers) are measured as the applied perpendicular magnetic field is increased from zero to $\approx$0.5~mT, which corresponds to a local field at the SQUID of $\approx$60~mT for chip~A and $\approx$20~mT for chip~B due to different flux focusing factors. The smaller size of chip~B likely contributes to the smaller flux focusing factor.

The resonance responses and their fits at these fields are shown in Fig.~\ref{fig:resfit}b, c. Even at these perpendicular magnetic fields, the internal quality factor of ATune (BTune) is $Q_{\rm i}=9.1\times$10$^3$ ($9.98\times 10^4$) and the resonator remains tunable, demonstrating the field resilience of these devices. Indeed Fig.~\ref{fig:large_field}c shows that BTune retains a high quality factor right across the field range shown, even after tuning through 42 periods.

For ABare, $\nu_0$ and $Q_{\rm i}$ tune weakly and smoothly as the applied magnetic field is increased up to about 0.21~mT with the exception of abrupt drops in both $\nu_0$ and $Q_{\rm i}$ at four field values up to 0.2~mT (see Fig.~\ref{fig:large_field}). 
This step-like response of $Q_{\rm i}$ vs magnetic field is consistent with the formation of vortices on the superconducting resonator's central conductor~\cite{nsanzinea2014}. 
For ATune, $Q_{\rm i}$ modulates by a factor of about 5 with field as the resonator tunes (as previously shown in Fig.~\ref{fig:up_down_single_period}). In addition to the modulation with tuning, the maximum  $Q_{\rm i}$ also drops with increasing applied field. The magnitude and field scale of this drop are similar to those of the bare resonator.
The respective $Q_{\rm i}(B)$ dependences for ATune and ABare are consistent with the maximum of the field-modulated $Q_{\rm i}$ in ATune being limited, as $B$ increases, by the same physics which causes the step-like reductions in $Q_{\rm i}$ for ABare. The similarity is even more clearly seen in $\nu_0(B)$, where the untuned resonant frequency of ATune jumps up at an applied field of 0.21~mT just as in the bare resonator. This suggests that not only does the nanoSQUID introduce minimal loss at zero applied field, the \textit{resonator's} field resilience is the limiting factor in the field performance of the device. This is also consistent with measurements of nanoSQUIDs fabricated from ultrathin niobium films successfully operating in parallel fields up to 7~T~\cite{chen2010chip}. Our results are therefore promising for future generations of devices where SQUIDs are embedded within field-resilient resonators. Importantly, ATune operates at local fields comfortably above 30~mT (and BTune looks likely to also operate at such fields), the first clock transition of bismuth spins.

In the literature, resonator tuning is typically given in units of flux quanta. This means that there are no comparable results on field resilience of tunable resonators to which this device can be compared. Therefore, we believe that this study is the first report addressing the field resilience of tunable resonators.

\section{Conclusions}

In conclusion, we have embedded constriction nanoSQUIDs in Nb resonators, and shown that this device geometry results in a frequency tunable resonator with 100 MHz of tuning demonstrated in a 7.42 GHz resonator.
We fabricate these devices from two Nb chips and show that the untuned quality factor of these resonators is limited by the bare resonator technology rather than embedding the SQUID, up to internal quality factors of $Q_{\rm i} = 1.2\times10^5$ --- the highest quality factor SQUID-tunable resonator reported to date. 
These results are realised in two different resonators where one has high quality and the other is more tunable, however there is no fundamental trade-off being made between quality and tunability in these devices. By comparing the quality factors of tunable and bare resonators we demonstrate that the quality is determined by the superconducting films. The tunability differs due to different device geometries.
We show that even these high quality factors are resonator-limited both at zero-field and applied perpendicular magnetic fields up to 0.5~mT and for resonator photon number from  $10^4$ down to single photons at zero field. These resonators remain tunable at applied perpendicular fields of 0.5~mT and retain an untuned quality factor $\sim1\times10^5$. The devices presented here compare favourably with the highest quality tunable resonators previously reported  \cite{vissers2015frequency}. Those resonators tune due to kinetic inductance shifts from DC currents and achieve high-power quality factors of $1.8\times10^5$; there are no reports of low power operation.

A number of modifications to the device and measurement setup may straightforwardly be made to improve field resilience, specifically: operating the device in parallel fields (which tune spins to the clock transition without applying large fields to the resonator), modifying the device design by the addition of anti-dots~\cite{bothner2011improving}, patterning the whole ground-plane~\cite{graaf2012magnetic} and/or the use of resonator designs inherently more robust to external field~\cite{samkharadze2016high,bothner2011improving,graaf2012magnetic}. The nanoSQUIDs allow us to measure the local magnetic field around the resonators and we find that the local fields are 30--120 time greater than the applied fields meaning these modifications should straightforwardly significantly improve field resilience.

These resonators hold great promise for future hybrid quantum system applications. For example, a tunable resonator operating at 27mT coupled to Bi spins in Si could address spins at their clock transition. The resonator could be tuned into resonance with the spins, and ESR pulse sequences applied, before the resonator is tuned away from the spins in frequency, allowing the long-coherence-time spins to store quantum information, unperturbed by the resonators.

\section{Acknowledgements}
The authors thank E. J. Romans, A. Blois, S. Probst, P. Bertet, C. W. Zollitsch for useful discussions. This project has received funding from the European Union’s Horizon 2020 research and innovation programme under the Marie Sklodowska-Curie grant agreement No 705713 QUINTESSENS (E.~D.-F.), the European Community's Seventh Framework Programme through grant agreement No.~279781 (ASCENT) (J.M.), Carl Zeiss Semiconductor Manufacturing (O.~W.~K.) and the U.K. Engineering and Physical Sciences Research Council, Grant Numbers EP/J017329/1 (J.~C.~F.), EP/K024701/1 and EP/H005544/1 (O.~W.~K. and P.~A.~W.), EP/P510270/1 (O.~W.~K).

\bibliography{bibliography}

\begin{thebibliography}{42}%
\makeatletter
\providecommand \@ifxundefined [1]{%
 \@ifx{#1\undefined}
}%
\providecommand \@ifnum [1]{%
 \ifnum #1\expandafter \@firstoftwo
 \else \expandafter \@secondoftwo
 \fi
}%
\providecommand \@ifx [1]{%
 \ifx #1\expandafter \@firstoftwo
 \else \expandafter \@secondoftwo
 \fi
}%
\providecommand \natexlab [1]{#1}%
\providecommand \enquote  [1]{#1}%
\providecommand \bibnamefont  [1]{#1}%
\providecommand \bibfnamefont [1]{#1}%
\providecommand \citenamefont [1]{#1}%
\providecommand \href@noop [0]{\@secondoftwo}%
\providecommand \href [0]{\begingroup \@sanitize@url \@href}%
\providecommand \@href[1]{\@@startlink{#1}\@@href}%
\providecommand \@@href[1]{\endgroup#1\@@endlink}%
\providecommand \@sanitize@url [0]{\catcode `\\12\catcode `\$12\catcode
  `\&12\catcode `\#12\catcode `\^12\catcode `\_12\catcode `\%12\relax}%
\providecommand \@@startlink[1]{}%
\providecommand \@@endlink[0]{}%
\providecommand \url  [0]{\begingroup\@sanitize@url \@url }%
\providecommand \@url [1]{\endgroup\@href {#1}{\urlprefix }}%
\providecommand \urlprefix  [0]{URL }%
\providecommand \Eprint [0]{\href }%
\providecommand \doibase [0]{http://dx.doi.org/}%
\providecommand \selectlanguage [0]{\@gobble}%
\providecommand \bibinfo  [0]{\@secondoftwo}%
\providecommand \bibfield  [0]{\@secondoftwo}%
\providecommand \translation [1]{[#1]}%
\providecommand \BibitemOpen [0]{}%
\providecommand \bibitemStop [0]{}%
\providecommand \bibitemNoStop [0]{.\EOS\space}%
\providecommand \EOS [0]{\spacefactor3000\relax}%
\providecommand \BibitemShut  [1]{\csname bibitem#1\endcsname}%
\let\auto@bib@innerbib\@empty
\bibitem [{\citenamefont {Bienfait}\ \emph {et~al.}(2016)\citenamefont
  {Bienfait}, \citenamefont {Pla}, \citenamefont {Kubo}, \citenamefont {Zhou},
  \citenamefont {Stern}, \citenamefont {Lo}, \citenamefont {Weis},
  \citenamefont {Schenkel}, \citenamefont {Vion}, \citenamefont {Esteve},
  \citenamefont {Morton},\ and\ \citenamefont
  {Bertet}}]{bienfait2016controlling}%
  \BibitemOpen
  \bibfield  {author} {\bibinfo {author} {\bibfnamefont {A}~\bibnamefont
  {Bienfait}}, \bibinfo {author} {\bibfnamefont {JJ}~\bibnamefont {Pla}},
  \bibinfo {author} {\bibfnamefont {Y}~\bibnamefont {Kubo}}, \bibinfo {author}
  {\bibfnamefont {X}~\bibnamefont {Zhou}}, \bibinfo {author} {\bibfnamefont
  {M}~\bibnamefont {Stern}}, \bibinfo {author} {\bibfnamefont {CC}~\bibnamefont
  {Lo}}, \bibinfo {author} {\bibfnamefont {CD}~\bibnamefont {Weis}}, \bibinfo
  {author} {\bibfnamefont {T}~\bibnamefont {Schenkel}}, \bibinfo {author}
  {\bibfnamefont {D}~\bibnamefont {Vion}}, \bibinfo {author} {\bibfnamefont
  {D}~\bibnamefont {Esteve}}, \bibinfo {author} {\bibfnamefont {JJL}\
  \bibnamefont {Morton}}, \ and\ \bibinfo {author} {\bibfnamefont
  {P}~\bibnamefont {Bertet}},\ }\bibfield  {title} {\enquote {\bibinfo {title}
  {Controlling spin relaxation with a cavity},}\ }\href@noop {} {\bibfield
  {journal} {\bibinfo  {journal} {Nature}\ }\textbf {\bibinfo {volume} {531}},\
  \bibinfo {pages} {74} (\bibinfo {year} {2016})}\BibitemShut {NoStop}%
\bibitem [{\citenamefont {Kubo}\ \emph {et~al.}(2010)\citenamefont {Kubo},
  \citenamefont {Ong}, \citenamefont {Bertet}, \citenamefont {Vion},
  \citenamefont {Jacques}, \citenamefont {Zheng}, \citenamefont {Dr{\'e}au},
  \citenamefont {Roch}, \citenamefont {Auff{\`e}ves}, \citenamefont {Jelezko},
  \citenamefont {J}, \citenamefont {Barthe}, \citenamefont {Bergonzo},\ and\
  \citenamefont {Esteve}}]{kubo2010strong}%
  \BibitemOpen
  \bibfield  {author} {\bibinfo {author} {\bibfnamefont {Y}~\bibnamefont
  {Kubo}}, \bibinfo {author} {\bibfnamefont {FR}~\bibnamefont {Ong}}, \bibinfo
  {author} {\bibfnamefont {Patrice}\ \bibnamefont {Bertet}}, \bibinfo {author}
  {\bibfnamefont {Denis}\ \bibnamefont {Vion}}, \bibinfo {author}
  {\bibfnamefont {V}~\bibnamefont {Jacques}}, \bibinfo {author} {\bibfnamefont
  {D}~\bibnamefont {Zheng}}, \bibinfo {author} {\bibfnamefont {A}~\bibnamefont
  {Dr{\'e}au}}, \bibinfo {author} {\bibfnamefont {J-F}\ \bibnamefont {Roch}},
  \bibinfo {author} {\bibfnamefont {Alexia}\ \bibnamefont {Auff{\`e}ves}},
  \bibinfo {author} {\bibfnamefont {Fedor}\ \bibnamefont {Jelezko}}, \bibinfo
  {author} {\bibfnamefont {Wrachtrup}\ \bibnamefont {J}}, \bibinfo {author}
  {\bibfnamefont {MF}~\bibnamefont {Barthe}}, \bibinfo {author} {\bibfnamefont
  {P}~\bibnamefont {Bergonzo}}, \ and\ \bibinfo {author} {\bibfnamefont
  {D}~\bibnamefont {Esteve}},\ }\bibfield  {title} {\enquote {\bibinfo {title}
  {Strong coupling of a spin ensemble to a superconducting resonator},}\
  }\href@noop {} {\bibfield  {journal} {\bibinfo  {journal} {Physical Review
  Letters}\ }\textbf {\bibinfo {volume} {105}},\ \bibinfo {pages} {140502}
  (\bibinfo {year} {2010})}\BibitemShut {NoStop}%
\bibitem [{\citenamefont {Probst}\ \emph {et~al.}(2013)\citenamefont {Probst},
  \citenamefont {Rotzinger}, \citenamefont {W{\"u}nsch}, \citenamefont {Jung},
  \citenamefont {Jerger}, \citenamefont {Siegel}, \citenamefont {Ustinov},\
  and\ \citenamefont {Bushev}}]{probst2013anisotropic}%
  \BibitemOpen
  \bibfield  {author} {\bibinfo {author} {\bibfnamefont {S}~\bibnamefont
  {Probst}}, \bibinfo {author} {\bibfnamefont {H}~\bibnamefont {Rotzinger}},
  \bibinfo {author} {\bibfnamefont {S}~\bibnamefont {W{\"u}nsch}}, \bibinfo
  {author} {\bibfnamefont {P}~\bibnamefont {Jung}}, \bibinfo {author}
  {\bibfnamefont {M}~\bibnamefont {Jerger}}, \bibinfo {author} {\bibfnamefont
  {M}~\bibnamefont {Siegel}}, \bibinfo {author} {\bibfnamefont
  {AV}~\bibnamefont {Ustinov}}, \ and\ \bibinfo {author} {\bibfnamefont
  {PA}~\bibnamefont {Bushev}},\ }\bibfield  {title} {\enquote {\bibinfo {title}
  {Anisotropic rare-earth spin ensemble strongly coupled to a superconducting
  resonator},}\ }\href@noop {} {\bibfield  {journal} {\bibinfo  {journal}
  {Physical Review Letters}\ }\textbf {\bibinfo {volume} {110}},\ \bibinfo
  {pages} {157001} (\bibinfo {year} {2013})}\BibitemShut {NoStop}%
\bibitem [{\citenamefont {Huebl}\ \emph {et~al.}(2013)\citenamefont {Huebl},
  \citenamefont {Zollitsch}, \citenamefont {Lotze}, \citenamefont {Hocke},
  \citenamefont {Greifenstein}, \citenamefont {Marx}, \citenamefont {Gross},\
  and\ \citenamefont {Goennenwein}}]{huebl2013high}%
  \BibitemOpen
  \bibfield  {author} {\bibinfo {author} {\bibfnamefont {H}~\bibnamefont
  {Huebl}}, \bibinfo {author} {\bibfnamefont {CW}~\bibnamefont {Zollitsch}},
  \bibinfo {author} {\bibfnamefont {J}~\bibnamefont {Lotze}}, \bibinfo {author}
  {\bibfnamefont {F}~\bibnamefont {Hocke}}, \bibinfo {author} {\bibfnamefont
  {M}~\bibnamefont {Greifenstein}}, \bibinfo {author} {\bibfnamefont
  {A}~\bibnamefont {Marx}}, \bibinfo {author} {\bibfnamefont {R}~\bibnamefont
  {Gross}}, \ and\ \bibinfo {author} {\bibfnamefont {STB}\ \bibnamefont
  {Goennenwein}},\ }\bibfield  {title} {\enquote {\bibinfo {title} {High
  cooperativity in coupled microwave resonator ferrimagnetic insulator
  hybrids},}\ }\href@noop {} {\bibfield  {journal} {\bibinfo  {journal}
  {Physical Review Letters}\ }\textbf {\bibinfo {volume} {111}},\ \bibinfo
  {pages} {127003} (\bibinfo {year} {2013})}\BibitemShut {NoStop}%
\bibitem [{\citenamefont {Abdurakhimov}\ \emph {et~al.}(2018)\citenamefont
  {Abdurakhimov}, \citenamefont {Khan}, \citenamefont {Panjwani}, \citenamefont
  {Breeze}, \citenamefont {Seki}, \citenamefont {Tokura}, \citenamefont
  {Morton},\ and\ \citenamefont {Kurebayashi}}]{abdurakhimov2018strong}%
  \BibitemOpen
  \bibfield  {author} {\bibinfo {author} {\bibfnamefont {LV}~\bibnamefont
  {Abdurakhimov}}, \bibinfo {author} {\bibfnamefont {S}~\bibnamefont {Khan}},
  \bibinfo {author} {\bibfnamefont {NA}~\bibnamefont {Panjwani}}, \bibinfo
  {author} {\bibfnamefont {JD}~\bibnamefont {Breeze}}, \bibinfo {author}
  {\bibfnamefont {S}~\bibnamefont {Seki}}, \bibinfo {author} {\bibfnamefont
  {Y}~\bibnamefont {Tokura}}, \bibinfo {author} {\bibfnamefont {JJL}\
  \bibnamefont {Morton}}, \ and\ \bibinfo {author} {\bibfnamefont
  {H}~\bibnamefont {Kurebayashi}},\ }\bibfield  {title} {\enquote {\bibinfo
  {title} {{Strong coupling between magnons in a chiral magnetic insulator
  ${\rm Cu_2OSeO_3}$ and microwave cavity photons}},}\ }\href@noop {}
  {\bibfield  {journal} {\bibinfo  {journal} {arXiv preprint arXiv:1802.07113}\
  } (\bibinfo {year} {2018})}\BibitemShut {NoStop}%
\bibitem [{\citenamefont {Mi}\ \emph {et~al.}(2017)\citenamefont {Mi},
  \citenamefont {Cady}, \citenamefont {Zajac}, \citenamefont {Deelman},\ and\
  \citenamefont {Petta}}]{mi2016strong}%
  \BibitemOpen
  \bibfield  {author} {\bibinfo {author} {\bibfnamefont {Xiao}\ \bibnamefont
  {Mi}}, \bibinfo {author} {\bibfnamefont {JV}~\bibnamefont {Cady}}, \bibinfo
  {author} {\bibfnamefont {DM}~\bibnamefont {Zajac}}, \bibinfo {author}
  {\bibfnamefont {PW}~\bibnamefont {Deelman}}, \ and\ \bibinfo {author}
  {\bibfnamefont {JR}~\bibnamefont {Petta}},\ }\bibfield  {title} {\enquote
  {\bibinfo {title} {Strong coupling of a single electron in silicon to a
  microwave photon},}\ }\href@noop {} {\bibfield  {journal} {\bibinfo
  {journal} {Science}\ }\textbf {\bibinfo {volume} {355}},\ \bibinfo {pages}
  {156--158} (\bibinfo {year} {2017})}\BibitemShut {NoStop}%
\bibitem [{\citenamefont {Gr{\`e}zes}\ \emph {et~al.}(2016)\citenamefont
  {Gr{\`e}zes}, \citenamefont {Kubo}, \citenamefont {Julsgaard}, \citenamefont
  {Umeda}, \citenamefont {Isoya}, \citenamefont {Sumiya}, \citenamefont {Abe},
  \citenamefont {Onoda}, \citenamefont {Ohshima}, \citenamefont {Nakamura},
  \citenamefont {Diniz}, \citenamefont {Auffeves}, \citenamefont {Jacques},
  \citenamefont {Roch}, \citenamefont {Vion}, \citenamefont {Esteve},
  \citenamefont {Moelmer},\ and\ \citenamefont {Bertet}}]{grezes2016towards}%
  \BibitemOpen
  \bibfield  {author} {\bibinfo {author} {\bibfnamefont {C}~\bibnamefont
  {Gr{\`e}zes}}, \bibinfo {author} {\bibfnamefont {Y}~\bibnamefont {Kubo}},
  \bibinfo {author} {\bibfnamefont {B}~\bibnamefont {Julsgaard}}, \bibinfo
  {author} {\bibfnamefont {T}~\bibnamefont {Umeda}}, \bibinfo {author}
  {\bibfnamefont {J}~\bibnamefont {Isoya}}, \bibinfo {author} {\bibfnamefont
  {H}~\bibnamefont {Sumiya}}, \bibinfo {author} {\bibfnamefont {H}~\bibnamefont
  {Abe}}, \bibinfo {author} {\bibfnamefont {S}~\bibnamefont {Onoda}}, \bibinfo
  {author} {\bibfnamefont {T}~\bibnamefont {Ohshima}}, \bibinfo {author}
  {\bibfnamefont {K}~\bibnamefont {Nakamura}}, \bibinfo {author} {\bibfnamefont
  {I}~\bibnamefont {Diniz}}, \bibinfo {author} {\bibfnamefont {A}~\bibnamefont
  {Auffeves}}, \bibinfo {author} {\bibfnamefont {V}~\bibnamefont {Jacques}},
  \bibinfo {author} {\bibfnamefont {J-F}\ \bibnamefont {Roch}}, \bibinfo
  {author} {\bibfnamefont {D}~\bibnamefont {Vion}}, \bibinfo {author}
  {\bibfnamefont {D}~\bibnamefont {Esteve}}, \bibinfo {author} {\bibfnamefont
  {K}~\bibnamefont {Moelmer}}, \ and\ \bibinfo {author} {\bibfnamefont
  {P}~\bibnamefont {Bertet}},\ }\bibfield  {title} {\enquote {\bibinfo {title}
  {Towards a spin-ensemble quantum memory for superconducting qubits},}\
  }\href@noop {} {\bibfield  {journal} {\bibinfo  {journal} {Comptes Rendus
  Physique}\ }\textbf {\bibinfo {volume} {17}},\ \bibinfo {pages} {693--704}
  (\bibinfo {year} {2016})}\BibitemShut {NoStop}%
\bibitem [{\citenamefont {Larsen}\ \emph {et~al.}(2015)\citenamefont {Larsen},
  \citenamefont {Petersson}, \citenamefont {Kuemmeth}, \citenamefont
  {Jespersen}, \citenamefont {Krogstrup}, \citenamefont {Nyg\aa{}rd},\ and\
  \citenamefont {Marcus}}]{Larsen2015}%
  \BibitemOpen
  \bibfield  {author} {\bibinfo {author} {\bibfnamefont {TW}~\bibnamefont
  {Larsen}}, \bibinfo {author} {\bibfnamefont {KD}~\bibnamefont {Petersson}},
  \bibinfo {author} {\bibfnamefont {F}~\bibnamefont {Kuemmeth}}, \bibinfo
  {author} {\bibfnamefont {TS}~\bibnamefont {Jespersen}}, \bibinfo {author}
  {\bibfnamefont {P}~\bibnamefont {Krogstrup}}, \bibinfo {author}
  {\bibfnamefont {J}~\bibnamefont {Nyg\aa{}rd}}, \ and\ \bibinfo {author}
  {\bibfnamefont {CM}~\bibnamefont {Marcus}},\ }\bibfield  {title} {\enquote
  {\bibinfo {title} {Semiconductor-nanowire-based superconducting qubit},}\
  }\href {\doibase 10.1103/PhysRevLett.115.127001} {\bibfield  {journal}
  {\bibinfo  {journal} {Phyiscal Review Letters}\ }\textbf {\bibinfo {volume}
  {115}},\ \bibinfo {pages} {127001} (\bibinfo {year} {2015})}\BibitemShut
  {NoStop}%
\bibitem [{\citenamefont {Yan}\ \emph {et~al.}(2016)\citenamefont {Yan},
  \citenamefont {Gustavsson}, \citenamefont {Kamal}, \citenamefont {Birenbaum},
  \citenamefont {Sears}, \citenamefont {Hover}, \citenamefont {Gudmundsen},
  \citenamefont {Rosenberg}, \citenamefont {Samach}, \citenamefont {Weber},
  \citenamefont {Yoder}, \citenamefont {Orlando}, \citenamefont {Clarke},
  \citenamefont {Kerman},\ and\ \citenamefont {Oliver}}]{yan2016flux}%
  \BibitemOpen
  \bibfield  {author} {\bibinfo {author} {\bibfnamefont {F}~\bibnamefont
  {Yan}}, \bibinfo {author} {\bibfnamefont {S}~\bibnamefont {Gustavsson}},
  \bibinfo {author} {\bibfnamefont {A}~\bibnamefont {Kamal}}, \bibinfo {author}
  {\bibfnamefont {J}~\bibnamefont {Birenbaum}}, \bibinfo {author}
  {\bibfnamefont {AP}~\bibnamefont {Sears}}, \bibinfo {author} {\bibfnamefont
  {D}~\bibnamefont {Hover}}, \bibinfo {author} {\bibfnamefont {TJ}~\bibnamefont
  {Gudmundsen}}, \bibinfo {author} {\bibfnamefont {D}~\bibnamefont
  {Rosenberg}}, \bibinfo {author} {\bibfnamefont {G}~\bibnamefont {Samach}},
  \bibinfo {author} {\bibfnamefont {S}~\bibnamefont {Weber}}, \bibinfo {author}
  {\bibfnamefont {JL}~\bibnamefont {Yoder}}, \bibinfo {author} {\bibfnamefont
  {TP}~\bibnamefont {Orlando}}, \bibinfo {author} {\bibfnamefont
  {J}~\bibnamefont {Clarke}}, \bibinfo {author} {\bibfnamefont
  {AJ}~\bibnamefont {Kerman}}, \ and\ \bibinfo {author} {\bibfnamefont
  {WD}~\bibnamefont {Oliver}},\ }\bibfield  {title} {\enquote {\bibinfo {title}
  {The flux qubit revisited to enhance coherence and reproducibility},}\
  }\href@noop {} {\bibfield  {journal} {\bibinfo  {journal} {Nature
  Communications}\ }\textbf {\bibinfo {volume} {7}},\ \bibinfo {pages} {12964}
  (\bibinfo {year} {2016})}\BibitemShut {NoStop}%
\bibitem [{\citenamefont {Dunsworth}\ \emph {et~al.}(2017)\citenamefont
  {Dunsworth}, \citenamefont {Megrant}, \citenamefont {Quintana}, \citenamefont
  {Chen}, \citenamefont {Barends}, \citenamefont {Burkett}, \citenamefont
  {Foxen}, \citenamefont {Chen}, \citenamefont {Chiaro}, \citenamefont
  {Fowler}, \citenamefont {Graff}, \citenamefont {Jeffrey}, \citenamefont
  {Kelly}, \citenamefont {Lucero}, \citenamefont {Mutus}, \citenamefont
  {Neeley}, \citenamefont {Neill}, \citenamefont {Roushan}, \citenamefont
  {Sank}, \citenamefont {Vainsencher}, \citenamefont {Wenner}, \citenamefont
  {White},\ and\ \citenamefont {Martinis}}]{Dunsworthloss}%
  \BibitemOpen
  \bibfield  {author} {\bibinfo {author} {\bibfnamefont {A}~\bibnamefont
  {Dunsworth}}, \bibinfo {author} {\bibfnamefont {A}~\bibnamefont {Megrant}},
  \bibinfo {author} {\bibfnamefont {C}~\bibnamefont {Quintana}}, \bibinfo
  {author} {\bibfnamefont {Z}~\bibnamefont {Chen}}, \bibinfo {author}
  {\bibfnamefont {R}~\bibnamefont {Barends}}, \bibinfo {author} {\bibfnamefont
  {B}~\bibnamefont {Burkett}}, \bibinfo {author} {\bibfnamefont
  {B}~\bibnamefont {Foxen}}, \bibinfo {author} {\bibfnamefont {Y}~\bibnamefont
  {Chen}}, \bibinfo {author} {\bibfnamefont {B}~\bibnamefont {Chiaro}},
  \bibinfo {author} {\bibfnamefont {A}~\bibnamefont {Fowler}}, \bibinfo
  {author} {\bibfnamefont {R}~\bibnamefont {Graff}}, \bibinfo {author}
  {\bibfnamefont {E}~\bibnamefont {Jeffrey}}, \bibinfo {author} {\bibfnamefont
  {J}~\bibnamefont {Kelly}}, \bibinfo {author} {\bibfnamefont {E}~\bibnamefont
  {Lucero}}, \bibinfo {author} {\bibfnamefont {JY}~\bibnamefont {Mutus}},
  \bibinfo {author} {\bibfnamefont {M}~\bibnamefont {Neeley}}, \bibinfo
  {author} {\bibfnamefont {C}~\bibnamefont {Neill}}, \bibinfo {author}
  {\bibfnamefont {P}~\bibnamefont {Roushan}}, \bibinfo {author} {\bibfnamefont
  {D}~\bibnamefont {Sank}}, \bibinfo {author} {\bibfnamefont {A}~\bibnamefont
  {Vainsencher}}, \bibinfo {author} {\bibfnamefont {J}~\bibnamefont {Wenner}},
  \bibinfo {author} {\bibfnamefont {TC}~\bibnamefont {White}}, \ and\ \bibinfo
  {author} {\bibfnamefont {JM}~\bibnamefont {Martinis}},\ }\bibfield  {title}
  {\enquote {\bibinfo {title} {Characterization and reduction of capacitive
  loss induced by sub-micron josephson junction fabrication in superconducting
  qubits},}\ }\href@noop {} {\bibfield  {journal} {\bibinfo  {journal} {Applied
  Physics Letters}\ }\textbf {\bibinfo {volume} {111}},\ \bibinfo {pages}
  {022601} (\bibinfo {year} {2017})}\BibitemShut {NoStop}%
\bibitem [{\citenamefont {Blum}\ \emph {et~al.}(2015)\citenamefont {Blum},
  \citenamefont {O'Brien}, \citenamefont {Lauk}, \citenamefont {Bushev},
  \citenamefont {Fleischhauer},\ and\ \citenamefont {Morigi}}]{Blum2015}%
  \BibitemOpen
  \bibfield  {author} {\bibinfo {author} {\bibfnamefont {S}~\bibnamefont
  {Blum}}, \bibinfo {author} {\bibfnamefont {C}~\bibnamefont {O'Brien}},
  \bibinfo {author} {\bibfnamefont {N}~\bibnamefont {Lauk}}, \bibinfo {author}
  {\bibfnamefont {P}~\bibnamefont {Bushev}}, \bibinfo {author} {\bibfnamefont
  {M}~\bibnamefont {Fleischhauer}}, \ and\ \bibinfo {author} {\bibfnamefont
  {G}~\bibnamefont {Morigi}},\ }\bibfield  {title} {\enquote {\bibinfo {title}
  {Interfacing microwave qubits and optical photons via spin ensembles},}\
  }\href {\doibase 10.1103/PhysRevA.91.033834} {\bibfield  {journal} {\bibinfo
  {journal} {Phyiscal Review A}\ }\textbf {\bibinfo {volume} {91}},\ \bibinfo
  {pages} {033834} (\bibinfo {year} {2015})}\BibitemShut {NoStop}%
\bibitem [{\citenamefont {Burnett}\ \emph {et~al.}(2014)\citenamefont
  {Burnett}, \citenamefont {Faoro}, \citenamefont {Wisby}, \citenamefont
  {Gurtovoi}, \citenamefont {Chernykh}, \citenamefont {Mikhailov},
  \citenamefont {Tulin}, \citenamefont {Shaikhaidarov}, \citenamefont
  {Antonov}, \citenamefont {Meeson}, \citenamefont {Tzalenchuk},\ and\
  \citenamefont {Lindstr{\"o}m}}]{burnett2014evidence}%
  \BibitemOpen
  \bibfield  {author} {\bibinfo {author} {\bibfnamefont {J}~\bibnamefont
  {Burnett}}, \bibinfo {author} {\bibfnamefont {L}~\bibnamefont {Faoro}},
  \bibinfo {author} {\bibfnamefont {I}~\bibnamefont {Wisby}}, \bibinfo {author}
  {\bibfnamefont {VL}~\bibnamefont {Gurtovoi}}, \bibinfo {author}
  {\bibfnamefont {AV}~\bibnamefont {Chernykh}}, \bibinfo {author}
  {\bibfnamefont {GM}~\bibnamefont {Mikhailov}}, \bibinfo {author}
  {\bibfnamefont {VA}~\bibnamefont {Tulin}}, \bibinfo {author} {\bibfnamefont
  {R}~\bibnamefont {Shaikhaidarov}}, \bibinfo {author} {\bibfnamefont
  {V}~\bibnamefont {Antonov}}, \bibinfo {author} {\bibfnamefont
  {PJ}~\bibnamefont {Meeson}}, \bibinfo {author} {\bibfnamefont
  {AY}~\bibnamefont {Tzalenchuk}}, \ and\ \bibinfo {author} {\bibfnamefont
  {T}~\bibnamefont {Lindstr{\"o}m}},\ }\bibfield  {title} {\enquote {\bibinfo
  {title} {Evidence for interacting two-level systems from the 1/f noise of a
  superconducting resonator},}\ }\href@noop {} {\bibfield  {journal} {\bibinfo
  {journal} {Nature Communications}\ }\textbf {\bibinfo {volume} {5}},\
  \bibinfo {pages} {4119} (\bibinfo {year} {2014})}\BibitemShut {NoStop}%
\bibitem [{\citenamefont {Barends}\ \emph {et~al.}(2014)\citenamefont
  {Barends}, \citenamefont {Kelly}, \citenamefont {Megrant}, \citenamefont
  {Veitia}, \citenamefont {Sank}, \citenamefont {Jeffrey}, \citenamefont
  {White}, \citenamefont {Mutus}, \citenamefont {Fowler}, \citenamefont
  {Campbell}, \citenamefont {Chen}, \citenamefont {Chen}, \citenamefont
  {Chiaro}, \citenamefont {Dunsworth}, \citenamefont {Neill}, \citenamefont
  {O'Malley}, \citenamefont {Roushan}, \citenamefont {Vainsencher},
  \citenamefont {Wenner}, \citenamefont {Korotkov}, \citenamefont {AN},\ and\
  \citenamefont {Martinis}}]{barends2014superconducting}%
  \BibitemOpen
  \bibfield  {author} {\bibinfo {author} {\bibfnamefont {R}~\bibnamefont
  {Barends}}, \bibinfo {author} {\bibfnamefont {J}~\bibnamefont {Kelly}},
  \bibinfo {author} {\bibfnamefont {A}~\bibnamefont {Megrant}}, \bibinfo
  {author} {\bibfnamefont {A}~\bibnamefont {Veitia}}, \bibinfo {author}
  {\bibfnamefont {D}~\bibnamefont {Sank}}, \bibinfo {author} {\bibfnamefont
  {E}~\bibnamefont {Jeffrey}}, \bibinfo {author} {\bibfnamefont
  {TC}~\bibnamefont {White}}, \bibinfo {author} {\bibfnamefont {J}~\bibnamefont
  {Mutus}}, \bibinfo {author} {\bibfnamefont {AG}~\bibnamefont {Fowler}},
  \bibinfo {author} {\bibfnamefont {B}~\bibnamefont {Campbell}}, \bibinfo
  {author} {\bibfnamefont {Y}~\bibnamefont {Chen}}, \bibinfo {author}
  {\bibfnamefont {Z}~\bibnamefont {Chen}}, \bibinfo {author} {\bibfnamefont
  {B}~\bibnamefont {Chiaro}}, \bibinfo {author} {\bibfnamefont {A}~\bibnamefont
  {Dunsworth}}, \bibinfo {author} {\bibfnamefont {C}~\bibnamefont {Neill}},
  \bibinfo {author} {\bibfnamefont {P}~\bibnamefont {O'Malley}}, \bibinfo
  {author} {\bibfnamefont {P}~\bibnamefont {Roushan}}, \bibinfo {author}
  {\bibfnamefont {A}~\bibnamefont {Vainsencher}}, \bibinfo {author}
  {\bibfnamefont {J}~\bibnamefont {Wenner}}, \bibinfo {author} {\bibfnamefont
  {AN}~\bibnamefont {Korotkov}}, \bibinfo {author} {\bibfnamefont {Cleland}\
  \bibnamefont {AN}}, \ and\ \bibinfo {author} {\bibfnamefont {JM}~\bibnamefont
  {Martinis}},\ }\bibfield  {title} {\enquote {\bibinfo {title}
  {Superconducting quantum circuits at the surface code threshold for fault
  tolerance},}\ }\href@noop {} {\bibfield  {journal} {\bibinfo  {journal}
  {Nature}\ }\textbf {\bibinfo {volume} {508}},\ \bibinfo {pages} {500--503}
  (\bibinfo {year} {2014})}\BibitemShut {NoStop}%
\bibitem [{\citenamefont {Megrant}\ \emph {et~al.}(2012)\citenamefont
  {Megrant}, \citenamefont {Neill}, \citenamefont {Barends}, \citenamefont
  {Chiaro}, \citenamefont {Chen}, \citenamefont {Feigl}, \citenamefont {Kelly},
  \citenamefont {Lucero}, \citenamefont {Mariantoni}, \citenamefont {O'Malley},
  \citenamefont {Sank}, \citenamefont {Vainsencher}, \citenamefont {Wenner},
  \citenamefont {White}, \citenamefont {Yin}, \citenamefont {Zhao},
  \citenamefont {CJ}, \citenamefont {Martinis},\ and\ \citenamefont
  {Cleland}}]{megrant2012planar}%
  \BibitemOpen
  \bibfield  {author} {\bibinfo {author} {\bibfnamefont {A}~\bibnamefont
  {Megrant}}, \bibinfo {author} {\bibfnamefont {C}~\bibnamefont {Neill}},
  \bibinfo {author} {\bibfnamefont {R}~\bibnamefont {Barends}}, \bibinfo
  {author} {\bibfnamefont {B}~\bibnamefont {Chiaro}}, \bibinfo {author}
  {\bibfnamefont {Yu}~\bibnamefont {Chen}}, \bibinfo {author} {\bibfnamefont
  {L}~\bibnamefont {Feigl}}, \bibinfo {author} {\bibfnamefont {J}~\bibnamefont
  {Kelly}}, \bibinfo {author} {\bibfnamefont {Erik}\ \bibnamefont {Lucero}},
  \bibinfo {author} {\bibfnamefont {Matteo}\ \bibnamefont {Mariantoni}},
  \bibinfo {author} {\bibfnamefont {PJJ}\ \bibnamefont {O'Malley}}, \bibinfo
  {author} {\bibfnamefont {D}~\bibnamefont {Sank}}, \bibinfo {author}
  {\bibfnamefont {A}~\bibnamefont {Vainsencher}}, \bibinfo {author}
  {\bibfnamefont {J}~\bibnamefont {Wenner}}, \bibinfo {author} {\bibfnamefont
  {TC}~\bibnamefont {White}}, \bibinfo {author} {\bibfnamefont {Y}~\bibnamefont
  {Yin}}, \bibinfo {author} {\bibfnamefont {J}~\bibnamefont {Zhao}}, \bibinfo
  {author} {\bibfnamefont {Palmstr{\o}m}\ \bibnamefont {CJ}}, \bibinfo {author}
  {\bibfnamefont {John~M}\ \bibnamefont {Martinis}}, \ and\ \bibinfo {author}
  {\bibfnamefont {AN}~\bibnamefont {Cleland}},\ }\bibfield  {title} {\enquote
  {\bibinfo {title} {Planar superconducting resonators with internal quality
  factors above one million},}\ }\href@noop {} {\bibfield  {journal} {\bibinfo
  {journal} {Applied Physics Letters}\ }\textbf {\bibinfo {volume} {100}},\
  \bibinfo {pages} {113510} (\bibinfo {year} {2012})}\BibitemShut {NoStop}%
\bibitem [{\citenamefont {Palacios-Laloy}\ \emph {et~al.}(2008)\citenamefont
  {Palacios-Laloy}, \citenamefont {Nguyen}, \citenamefont {Mallet},
  \citenamefont {Bertet}, \citenamefont {Vion},\ and\ \citenamefont
  {Esteve}}]{palacios2008tunable}%
  \BibitemOpen
  \bibfield  {author} {\bibinfo {author} {\bibfnamefont {A}~\bibnamefont
  {Palacios-Laloy}}, \bibinfo {author} {\bibfnamefont {F}~\bibnamefont
  {Nguyen}}, \bibinfo {author} {\bibfnamefont {F}~\bibnamefont {Mallet}},
  \bibinfo {author} {\bibfnamefont {P}~\bibnamefont {Bertet}}, \bibinfo
  {author} {\bibfnamefont {D}~\bibnamefont {Vion}}, \ and\ \bibinfo {author}
  {\bibfnamefont {D}~\bibnamefont {Esteve}},\ }\bibfield  {title} {\enquote
  {\bibinfo {title} {Tunable resonators for quantum circuits},}\ }\href@noop {}
  {\bibfield  {journal} {\bibinfo  {journal} {Journal of Low Temperature
  Physics}\ }\textbf {\bibinfo {volume} {151}},\ \bibinfo {pages} {1034--1042}
  (\bibinfo {year} {2008})}\BibitemShut {NoStop}%
\bibitem [{\citenamefont {Sandberg}\ \emph {et~al.}(2008)\citenamefont
  {Sandberg}, \citenamefont {Wilson}, \citenamefont {Persson}, \citenamefont
  {Bauch}, \citenamefont {Johansson}, \citenamefont {Shumeiko}, \citenamefont
  {Duty},\ and\ \citenamefont {Delsing}}]{sandberg2008tuning}%
  \BibitemOpen
  \bibfield  {author} {\bibinfo {author} {\bibfnamefont {M}~\bibnamefont
  {Sandberg}}, \bibinfo {author} {\bibfnamefont {CM}~\bibnamefont {Wilson}},
  \bibinfo {author} {\bibfnamefont {F}~\bibnamefont {Persson}}, \bibinfo
  {author} {\bibfnamefont {T}~\bibnamefont {Bauch}}, \bibinfo {author}
  {\bibfnamefont {G}~\bibnamefont {Johansson}}, \bibinfo {author}
  {\bibfnamefont {V}~\bibnamefont {Shumeiko}}, \bibinfo {author} {\bibfnamefont
  {T}~\bibnamefont {Duty}}, \ and\ \bibinfo {author} {\bibfnamefont
  {P}~\bibnamefont {Delsing}},\ }\bibfield  {title} {\enquote {\bibinfo {title}
  {Tuning the field in a microwave resonator faster than the photon
  lifetime},}\ }\href@noop {} {\bibfield  {journal} {\bibinfo  {journal}
  {Applied Physics Letters}\ }\textbf {\bibinfo {volume} {92}},\ \bibinfo
  {pages} {203501} (\bibinfo {year} {2008})}\BibitemShut {NoStop}%
\bibitem [{\citenamefont {Levenson-Falk}\ \emph {et~al.}(2011)\citenamefont
  {Levenson-Falk}, \citenamefont {Vijay},\ and\ \citenamefont
  {Siddiqi}}]{levenson2011nonlinear}%
  \BibitemOpen
  \bibfield  {author} {\bibinfo {author} {\bibfnamefont {EM}~\bibnamefont
  {Levenson-Falk}}, \bibinfo {author} {\bibfnamefont {R}~\bibnamefont {Vijay}},
  \ and\ \bibinfo {author} {\bibfnamefont {I}~\bibnamefont {Siddiqi}},\
  }\bibfield  {title} {\enquote {\bibinfo {title} {Nonlinear microwave response
  of aluminum weak-link josephson oscillators},}\ }\href@noop {} {\bibfield
  {journal} {\bibinfo  {journal} {Applied Physics Letters}\ }\textbf {\bibinfo
  {volume} {98}},\ \bibinfo {pages} {3115} (\bibinfo {year}
  {2011})}\BibitemShut {NoStop}%
\bibitem [{\citenamefont {Adamyan}\ \emph {et~al.}(2016)\citenamefont
  {Adamyan}, \citenamefont {Kubatkin},\ and\ \citenamefont
  {Danilov}}]{adamyan2016tunable}%
  \BibitemOpen
  \bibfield  {author} {\bibinfo {author} {\bibfnamefont {AA}~\bibnamefont
  {Adamyan}}, \bibinfo {author} {\bibfnamefont {SE}~\bibnamefont {Kubatkin}}, \
  and\ \bibinfo {author} {\bibfnamefont {AV}~\bibnamefont {Danilov}},\
  }\bibfield  {title} {\enquote {\bibinfo {title} {Tunable superconducting
  microstrip resonators},}\ }\href@noop {} {\bibfield  {journal} {\bibinfo
  {journal} {Applied Physics Letters}\ }\textbf {\bibinfo {volume} {108}},\
  \bibinfo {pages} {172601} (\bibinfo {year} {2016})}\BibitemShut {NoStop}%
\bibitem [{\citenamefont {Asfaw}\ \emph {et~al.}(2017)\citenamefont {Asfaw},
  \citenamefont {Sigillito}, \citenamefont {Tyryshkin}, \citenamefont
  {Schenkel},\ and\ \citenamefont {Lyon}}]{asfaw2017multi}%
  \BibitemOpen
  \bibfield  {author} {\bibinfo {author} {\bibfnamefont {AT}~\bibnamefont
  {Asfaw}}, \bibinfo {author} {\bibfnamefont {AJ}~\bibnamefont {Sigillito}},
  \bibinfo {author} {\bibfnamefont {AM}~\bibnamefont {Tyryshkin}}, \bibinfo
  {author} {\bibfnamefont {T}~\bibnamefont {Schenkel}}, \ and\ \bibinfo
  {author} {\bibfnamefont {SA}~\bibnamefont {Lyon}},\ }\bibfield  {title}
  {\enquote {\bibinfo {title} {Multi-frequency spin manipulation using rapidly
  tunable superconducting coplanar waveguide microresonators},}\ }\href@noop {}
  {\bibfield  {journal} {\bibinfo  {journal} {Applied Physics Letters}\
  }\textbf {\bibinfo {volume} {111}},\ \bibinfo {pages} {032601} (\bibinfo
  {year} {2017})}\BibitemShut {NoStop}%
\bibitem [{\citenamefont {Granata}\ and\ \citenamefont
  {Vettoliere}(2016)}]{granata2016nano}%
  \BibitemOpen
  \bibfield  {author} {\bibinfo {author} {\bibfnamefont {C}~\bibnamefont
  {Granata}}\ and\ \bibinfo {author} {\bibfnamefont {A}~\bibnamefont
  {Vettoliere}},\ }\bibfield  {title} {\enquote {\bibinfo {title} {{Nano
  superconducting quantum interference device: A powerful tool for nanoscale
  investigations}},}\ }\href@noop {} {\bibfield  {journal} {\bibinfo  {journal}
  {Physics Reports}\ }\textbf {\bibinfo {volume} {614}},\ \bibinfo {pages}
  {1--69} (\bibinfo {year} {2016})}\BibitemShut {NoStop}%
\bibitem [{\citenamefont {Mart{\'\i}nez-P{\'e}rez}\ \emph
  {et~al.}(2017)\citenamefont {Mart{\'\i}nez-P{\'e}rez}, \citenamefont
  {Kleiner},\ and\ \citenamefont {Koelle}}]{martineznanosquids}%
  \BibitemOpen
  \bibfield  {author} {\bibinfo {author} {\bibfnamefont {MJ}~\bibnamefont
  {Mart{\'\i}nez-P{\'e}rez}}, \bibinfo {author} {\bibfnamefont {R}~\bibnamefont
  {Kleiner}}, \ and\ \bibinfo {author} {\bibfnamefont {D}~\bibnamefont
  {Koelle}},\ }\bibfield  {title} {\enquote {\bibinfo {title} {{NanoSQUIDs
  Applied to the Investigation of Small Magnetic Systems}},}\ }in\ \href@noop
  {} {\emph {\bibinfo {booktitle} {The Oxford Handbook of Small
  Superconductors}}}\ (\bibinfo  {publisher} {OUP},\ \bibinfo {year}
  {2017})\BibitemShut {NoStop}%
\bibitem [{\citenamefont {Schwarz}\ \emph {et~al.}(2013)\citenamefont
  {Schwarz}, \citenamefont {Nagel}, \citenamefont {W{\"o}lbing}, \citenamefont
  {Kemmler}, \citenamefont {Kleiner},\ and\ \citenamefont
  {Koelle}}]{schwarz2013}%
  \BibitemOpen
  \bibfield  {author} {\bibinfo {author} {\bibfnamefont {T}~\bibnamefont
  {Schwarz}}, \bibinfo {author} {\bibfnamefont {J}~\bibnamefont {Nagel}},
  \bibinfo {author} {\bibfnamefont {R}~\bibnamefont {W{\"o}lbing}}, \bibinfo
  {author} {\bibfnamefont {M}~\bibnamefont {Kemmler}}, \bibinfo {author}
  {\bibfnamefont {R}~\bibnamefont {Kleiner}}, \ and\ \bibinfo {author}
  {\bibfnamefont {D}~\bibnamefont {Koelle}},\ }\bibfield  {title} {\enquote
  {\bibinfo {title} {Low-noise nano superconducting quantum interference device
  operating in tesla magnetic fields},}\ }\href@noop {} {\bibfield  {journal}
  {\bibinfo  {journal} {ACS Nano}\ }\textbf {\bibinfo {volume} {7}},\ \bibinfo
  {pages} {844--850} (\bibinfo {year} {2013})}\BibitemShut {NoStop}%
\bibitem [{\citenamefont {Hao}\ \emph {et~al.}(2009)\citenamefont {Hao},
  \citenamefont {Cox},\ and\ \citenamefont {Gallop}}]{hao2009characteristics}%
  \BibitemOpen
  \bibfield  {author} {\bibinfo {author} {\bibfnamefont {L}~\bibnamefont
  {Hao}}, \bibinfo {author} {\bibfnamefont {D~C}\ \bibnamefont {Cox}}, \ and\
  \bibinfo {author} {\bibfnamefont {J~C}\ \bibnamefont {Gallop}},\ }\bibfield
  {title} {\enquote {\bibinfo {title} {Characteristics of focused ion beam
  nanoscale josephson devices},}\ }\href
  {http://stacks.iop.org/0953-2048/22/i=6/a=064011} {\bibfield  {journal}
  {\bibinfo  {journal} {Superconductor Science and Technology}\ }\textbf
  {\bibinfo {volume} {22}},\ \bibinfo {pages} {064011} (\bibinfo {year}
  {2009})}\BibitemShut {NoStop}%
\bibitem [{\citenamefont {Jenkins}\ \emph {et~al.}(2014)\citenamefont
  {Jenkins}, \citenamefont {Naether}, \citenamefont {Ciria}, \citenamefont
  {Ses{\'e}}, \citenamefont {Atkinson}, \citenamefont {S{\'a}nchez-Azqueta},
  \citenamefont {del Barco}, \citenamefont {Majer}, \citenamefont {Zueco},\
  and\ \citenamefont {Luis}}]{jenkins2014nanoscale}%
  \BibitemOpen
  \bibfield  {author} {\bibinfo {author} {\bibfnamefont {MD}~\bibnamefont
  {Jenkins}}, \bibinfo {author} {\bibfnamefont {U}~\bibnamefont {Naether}},
  \bibinfo {author} {\bibfnamefont {M}~\bibnamefont {Ciria}}, \bibinfo {author}
  {\bibfnamefont {J}~\bibnamefont {Ses{\'e}}}, \bibinfo {author} {\bibfnamefont
  {J}~\bibnamefont {Atkinson}}, \bibinfo {author} {\bibfnamefont
  {C}~\bibnamefont {S{\'a}nchez-Azqueta}}, \bibinfo {author} {\bibfnamefont
  {E}~\bibnamefont {del Barco}}, \bibinfo {author} {\bibfnamefont
  {J}~\bibnamefont {Majer}}, \bibinfo {author} {\bibfnamefont {D}~\bibnamefont
  {Zueco}}, \ and\ \bibinfo {author} {\bibfnamefont {F}~\bibnamefont {Luis}},\
  }\bibfield  {title} {\enquote {\bibinfo {title} {Nanoscale constrictions in
  superconducting coplanar waveguide resonators},}\ }\href@noop {} {\bibfield
  {journal} {\bibinfo  {journal} {Applied Physics Letters}\ }\textbf {\bibinfo
  {volume} {105}},\ \bibinfo {pages} {162601} (\bibinfo {year}
  {2014})}\BibitemShut {NoStop}%
\bibitem [{\citenamefont {Burnett}\ \emph {et~al.}(2017)\citenamefont
  {Burnett}, \citenamefont {Sagar}, \citenamefont {Kennedy}, \citenamefont
  {Warburton},\ and\ \citenamefont {Fenton}}]{burnettnanowireloss}%
  \BibitemOpen
  \bibfield  {author} {\bibinfo {author} {\bibfnamefont {J}~\bibnamefont
  {Burnett}}, \bibinfo {author} {\bibfnamefont {J}~\bibnamefont {Sagar}},
  \bibinfo {author} {\bibfnamefont {OW}~\bibnamefont {Kennedy}}, \bibinfo
  {author} {\bibfnamefont {PA}~\bibnamefont {Warburton}}, \ and\ \bibinfo
  {author} {\bibfnamefont {JC}~\bibnamefont {Fenton}},\ }\bibfield  {title}
  {\enquote {\bibinfo {title} {Low-loss superconducting nanowire circuits using
  a neon focused ion beam},}\ }\href {\doibase 10.1103/PhysRevApplied.8.014039}
  {\bibfield  {journal} {\bibinfo  {journal} {Physical Review Applied}\
  }\textbf {\bibinfo {volume} {8}},\ \bibinfo {pages} {014039} (\bibinfo {year}
  {2017})}\BibitemShut {NoStop}%
\bibitem [{\citenamefont {Meservey}\ and\ \citenamefont
  {Schwartz}(1969)}]{meservey1969equilibrium}%
  \BibitemOpen
  \bibfield  {author} {\bibinfo {author} {\bibfnamefont {R}~\bibnamefont
  {Meservey}}\ and\ \bibinfo {author} {\bibfnamefont {BB}~\bibnamefont
  {Schwartz}},\ }\href@noop {} {\emph {\bibinfo {title} {{Equilibrium
  properties: Comparison of experimental results with predictions of the BCS
  theory}}}},\ \bibinfo {type} {Tech. Rep.}\ (\bibinfo  {institution}
  {Massachusetts Inst. of Tech., Cambridge},\ \bibinfo {year}
  {1969})\BibitemShut {NoStop}%
\bibitem [{SI()}]{SI}%
  \BibitemOpen
  \bibfield  {title} {\enquote {\bibinfo {title} {{See Supplemental Material at
  ... for details on further statistics on junction dimensions, measurements on
  additional devices and calculations of resonator and SQUID inductance.}}}\
  }\href@noop {} {\ }\BibitemShut {NoStop}%
\bibitem [{\citenamefont {Probst}\ \emph {et~al.}(2015)\citenamefont {Probst},
  \citenamefont {Song}, \citenamefont {Bushev}, \citenamefont {Ustinov},\ and\
  \citenamefont {Weides}}]{doi:10.1063/1.4907935}%
  \BibitemOpen
  \bibfield  {author} {\bibinfo {author} {\bibfnamefont {S}~\bibnamefont
  {Probst}}, \bibinfo {author} {\bibfnamefont {FB}~\bibnamefont {Song}},
  \bibinfo {author} {\bibfnamefont {PA}~\bibnamefont {Bushev}}, \bibinfo
  {author} {\bibfnamefont {AV}~\bibnamefont {Ustinov}}, \ and\ \bibinfo
  {author} {\bibfnamefont {M}~\bibnamefont {Weides}},\ }\bibfield  {title}
  {\enquote {\bibinfo {title} {Efficient and robust analysis of complex
  scattering data under noise in microwave resonators},}\ }\href@noop {}
  {\bibfield  {journal} {\bibinfo  {journal} {Review of Scientific
  Instruments}\ }\textbf {\bibinfo {volume} {86}},\ \bibinfo {pages} {024706}
  (\bibinfo {year} {2015})}\BibitemShut {NoStop}%
\bibitem [{Note1()}]{Note1}%
  \BibitemOpen
  \bibinfo {note} {Here, field has been changed in small (0.1~$\upmu $T) steps
  to approach maximal detuning whilst ensuring that the point of maximal
  detuning is not overshot before the sweep direction is reversed.}\BibitemShut
  {Stop}%
\bibitem [{\citenamefont {Golubov}\ \emph {et~al.}(2004)\citenamefont
  {Golubov}, \citenamefont {Kupriyanov},\ and\ \citenamefont
  {Il’Ichev}}]{golubov2004current}%
  \BibitemOpen
  \bibfield  {author} {\bibinfo {author} {\bibfnamefont {AA}~\bibnamefont
  {Golubov}}, \bibinfo {author} {\bibfnamefont {MY}~\bibnamefont {Kupriyanov}},
  \ and\ \bibinfo {author} {\bibfnamefont {E}~\bibnamefont {Il’Ichev}},\
  }\bibfield  {title} {\enquote {\bibinfo {title} {The current-phase relation
  in josephson junctions},}\ }\href@noop {} {\bibfield  {journal} {\bibinfo
  {journal} {Reviews of modern physics}\ }\textbf {\bibinfo {volume} {76}},\
  \bibinfo {pages} {411} (\bibinfo {year} {2004})}\BibitemShut {NoStop}%
\bibitem [{\citenamefont {Jones}\ \emph {et~al.}(2013)\citenamefont {Jones},
  \citenamefont {Huhtam{\"a}ki}, \citenamefont {Salmilehto}, \citenamefont
  {Tan},\ and\ \citenamefont {M{\"o}tt{\"o}nen}}]{jones2013tunable}%
  \BibitemOpen
  \bibfield  {author} {\bibinfo {author} {\bibfnamefont {PJ}~\bibnamefont
  {Jones}}, \bibinfo {author} {\bibfnamefont {JAM}\ \bibnamefont
  {Huhtam{\"a}ki}}, \bibinfo {author} {\bibfnamefont {J}~\bibnamefont
  {Salmilehto}}, \bibinfo {author} {\bibfnamefont {KY}~\bibnamefont {Tan}}, \
  and\ \bibinfo {author} {\bibfnamefont {M}~\bibnamefont {M{\"o}tt{\"o}nen}},\
  }\bibfield  {title} {\enquote {\bibinfo {title} {Tunable electromagnetic
  environment for superconducting quantum bits},}\ }\href@noop {} {\bibfield
  {journal} {\bibinfo  {journal} {Scientific Reports}\ }\textbf {\bibinfo
  {volume} {3}},\ \bibinfo {pages} {1987} (\bibinfo {year} {2013})}\BibitemShut
  {NoStop}%
\bibitem [{\citenamefont {de~Graaf}\ \emph {et~al.}(2017)\citenamefont
  {de~Graaf}, \citenamefont {Adamyan}, \citenamefont {Lindstr\"om},
  \citenamefont {Erts}, \citenamefont {Kubatkin}, \citenamefont {Tzalenchuk},\
  and\ \citenamefont {Danilov}}]{degraaf2017spin}%
  \BibitemOpen
  \bibfield  {author} {\bibinfo {author} {\bibfnamefont {SE}~\bibnamefont
  {de~Graaf}}, \bibinfo {author} {\bibfnamefont {AA}~\bibnamefont {Adamyan}},
  \bibinfo {author} {\bibfnamefont {T}~\bibnamefont {Lindstr\"om}}, \bibinfo
  {author} {\bibfnamefont {D}~\bibnamefont {Erts}}, \bibinfo {author}
  {\bibfnamefont {SE}~\bibnamefont {Kubatkin}}, \bibinfo {author}
  {\bibfnamefont {AYa}\ \bibnamefont {Tzalenchuk}}, \ and\ \bibinfo {author}
  {\bibfnamefont {AV}~\bibnamefont {Danilov}},\ }\bibfield  {title} {\enquote
  {\bibinfo {title} {{Direct Identification of Dilute Surface Spins on
  ${\mathrm{Al}}_{2}{\mathrm{O}}_{3}$: Origin of Flux Noise in Quantum
  Circuits}},}\ }\href {\doibase 10.1103/PhysRevLett.118.057703} {\bibfield
  {journal} {\bibinfo  {journal} {Phyiscal Review Letters}\ }\textbf {\bibinfo
  {volume} {118}},\ \bibinfo {pages} {057703} (\bibinfo {year}
  {2017})}\BibitemShut {NoStop}%
\bibitem [{\citenamefont {Ithier}\ \emph {et~al.}(2005)\citenamefont {Ithier},
  \citenamefont {Collin}, \citenamefont {Joyez}, \citenamefont {Meeson},
  \citenamefont {Vion}, \citenamefont {Esteve}, \citenamefont {Chiarello},
  \citenamefont {Shnirman}, \citenamefont {Makhlin}, \citenamefont {Schriefl},\
  and\ \citenamefont {Sch\"on}}]{ithier2005}%
  \BibitemOpen
  \bibfield  {author} {\bibinfo {author} {\bibfnamefont {G}~\bibnamefont
  {Ithier}}, \bibinfo {author} {\bibfnamefont {E}~\bibnamefont {Collin}},
  \bibinfo {author} {\bibfnamefont {P}~\bibnamefont {Joyez}}, \bibinfo {author}
  {\bibfnamefont {PJ}~\bibnamefont {Meeson}}, \bibinfo {author} {\bibfnamefont
  {D}~\bibnamefont {Vion}}, \bibinfo {author} {\bibfnamefont {D}~\bibnamefont
  {Esteve}}, \bibinfo {author} {\bibfnamefont {F}~\bibnamefont {Chiarello}},
  \bibinfo {author} {\bibfnamefont {A}~\bibnamefont {Shnirman}}, \bibinfo
  {author} {\bibfnamefont {Y}~\bibnamefont {Makhlin}}, \bibinfo {author}
  {\bibfnamefont {J}~\bibnamefont {Schriefl}}, \ and\ \bibinfo {author}
  {\bibfnamefont {G}~\bibnamefont {Sch\"on}},\ }\bibfield  {title} {\enquote
  {\bibinfo {title} {Decoherence in a superconducting quantum bit circuit},}\
  }\href {\doibase 10.1103/PhysRevB.72.134519} {\bibfield  {journal} {\bibinfo
  {journal} {Phyiscal Review B}\ }\textbf {\bibinfo {volume} {72}},\ \bibinfo
  {pages} {134519} (\bibinfo {year} {2005})}\BibitemShut {NoStop}%
\bibitem [{\citenamefont {Tinkham}(2004)}]{tinkham2004introduction}%
  \BibitemOpen
  \bibfield  {author} {\bibinfo {author} {\bibfnamefont {M}~\bibnamefont
  {Tinkham}},\ }\href@noop {} {\emph {\bibinfo {title} {Introduction to
  superconductivity}}}\ (\bibinfo  {publisher} {Courier Corporation},\ \bibinfo
  {year} {2004})\BibitemShut {NoStop}%
\bibitem [{\citenamefont {Bothner}\ \emph {et~al.}(2017)\citenamefont
  {Bothner}, \citenamefont {Wiedmaier}, \citenamefont {Ferdinand},
  \citenamefont {Kleiner},\ and\ \citenamefont
  {Koelle}}]{bothner2017improving}%
  \BibitemOpen
  \bibfield  {author} {\bibinfo {author} {\bibfnamefont {Daniel}\ \bibnamefont
  {Bothner}}, \bibinfo {author} {\bibfnamefont {Dominik}\ \bibnamefont
  {Wiedmaier}}, \bibinfo {author} {\bibfnamefont {Benedikt}\ \bibnamefont
  {Ferdinand}}, \bibinfo {author} {\bibfnamefont {Reinhold}\ \bibnamefont
  {Kleiner}}, \ and\ \bibinfo {author} {\bibfnamefont {Dieter}\ \bibnamefont
  {Koelle}},\ }\bibfield  {title} {\enquote {\bibinfo {title} {Improving
  superconducting resonators in magnetic fields by reduced field focussing and
  engineered flux screening},}\ }\href@noop {} {\bibfield  {journal} {\bibinfo
  {journal} {Physical Review Applied}\ }\textbf {\bibinfo {volume} {8}},\
  \bibinfo {pages} {034025} (\bibinfo {year} {2017})}\BibitemShut {NoStop}%
\bibitem [{\citenamefont {Kim}\ \emph {et~al.}(2015)\citenamefont {Kim},
  \citenamefont {Johansson},\ and\ \citenamefont {Nori}}]{kim2015circuit}%
  \BibitemOpen
  \bibfield  {author} {\bibinfo {author} {\bibfnamefont {Eun-jong}\
  \bibnamefont {Kim}}, \bibinfo {author} {\bibfnamefont {JR}~\bibnamefont
  {Johansson}}, \ and\ \bibinfo {author} {\bibfnamefont {Franco}\ \bibnamefont
  {Nori}},\ }\bibfield  {title} {\enquote {\bibinfo {title} {Circuit analog of
  quadratic optomechanics},}\ }\href@noop {} {\bibfield  {journal} {\bibinfo
  {journal} {Physical Review A}\ }\textbf {\bibinfo {volume} {91}},\ \bibinfo
  {pages} {033835} (\bibinfo {year} {2015})}\BibitemShut {NoStop}%
\bibitem [{\citenamefont {Nsanzineza}\ and\ \citenamefont
  {Plourde}(2014)}]{nsanzinea2014}%
  \BibitemOpen
  \bibfield  {author} {\bibinfo {author} {\bibfnamefont {I.}~\bibnamefont
  {Nsanzineza}}\ and\ \bibinfo {author} {\bibfnamefont {B.~L.~T.}\ \bibnamefont
  {Plourde}},\ }\bibfield  {title} {\enquote {\bibinfo {title} {Trapping a
  single vortex and reducing quasiparticles in a superconducting resonator},}\
  }\href {\doibase 10.1103/PhysRevLett.113.117002} {\bibfield  {journal}
  {\bibinfo  {journal} {Phyiscal Review Letters}\ }\textbf {\bibinfo {volume}
  {113}},\ \bibinfo {pages} {117002} (\bibinfo {year} {2014})}\BibitemShut
  {NoStop}%
\bibitem [{\citenamefont {Chen}\ \emph {et~al.}(2010)\citenamefont {Chen},
  \citenamefont {Wernsdorfer}, \citenamefont {Lampropoulos}, \citenamefont
  {Christou},\ and\ \citenamefont {Chiorescu}}]{chen2010chip}%
  \BibitemOpen
  \bibfield  {author} {\bibinfo {author} {\bibfnamefont {Lei}\ \bibnamefont
  {Chen}}, \bibinfo {author} {\bibfnamefont {Wolfgang}\ \bibnamefont
  {Wernsdorfer}}, \bibinfo {author} {\bibfnamefont {Christos}\ \bibnamefont
  {Lampropoulos}}, \bibinfo {author} {\bibfnamefont {George}\ \bibnamefont
  {Christou}}, \ and\ \bibinfo {author} {\bibfnamefont {Irinel}\ \bibnamefont
  {Chiorescu}},\ }\bibfield  {title} {\enquote {\bibinfo {title} {{On-chip
  SQUID measurements in the presence of high magnetic fields}},}\ }\href@noop
  {} {\bibfield  {journal} {\bibinfo  {journal} {Nanotechnology}\ }\textbf
  {\bibinfo {volume} {21}},\ \bibinfo {pages} {405504} (\bibinfo {year}
  {2010})}\BibitemShut {NoStop}%
\bibitem [{\citenamefont {Vissers}\ \emph {et~al.}(2015)\citenamefont
  {Vissers}, \citenamefont {Hubmayr}, \citenamefont {Sandberg}, \citenamefont
  {Chaudhuri}, \citenamefont {Bockstiegel},\ and\ \citenamefont
  {Gao}}]{vissers2015frequency}%
  \BibitemOpen
  \bibfield  {author} {\bibinfo {author} {\bibfnamefont {Michael~R}\
  \bibnamefont {Vissers}}, \bibinfo {author} {\bibfnamefont {Johannes}\
  \bibnamefont {Hubmayr}}, \bibinfo {author} {\bibfnamefont {Martin}\
  \bibnamefont {Sandberg}}, \bibinfo {author} {\bibfnamefont {Saptarshi}\
  \bibnamefont {Chaudhuri}}, \bibinfo {author} {\bibfnamefont {Clint}\
  \bibnamefont {Bockstiegel}}, \ and\ \bibinfo {author} {\bibfnamefont
  {Jiansong}\ \bibnamefont {Gao}},\ }\bibfield  {title} {\enquote {\bibinfo
  {title} {Frequency-tunable superconducting resonators via nonlinear kinetic
  inductance},}\ }\href@noop {} {\bibfield  {journal} {\bibinfo  {journal}
  {Applied Physics Letters}\ }\textbf {\bibinfo {volume} {107}},\ \bibinfo
  {pages} {062601} (\bibinfo {year} {2015})}\BibitemShut {NoStop}%
\bibitem [{\citenamefont {Bothner}\ \emph {et~al.}(2011)\citenamefont
  {Bothner}, \citenamefont {Gaber}, \citenamefont {Kemmler}, \citenamefont
  {Koelle},\ and\ \citenamefont {Kleiner}}]{bothner2011improving}%
  \BibitemOpen
  \bibfield  {author} {\bibinfo {author} {\bibfnamefont {D}~\bibnamefont
  {Bothner}}, \bibinfo {author} {\bibfnamefont {T}~\bibnamefont {Gaber}},
  \bibinfo {author} {\bibfnamefont {M}~\bibnamefont {Kemmler}}, \bibinfo
  {author} {\bibfnamefont {D}~\bibnamefont {Koelle}}, \ and\ \bibinfo {author}
  {\bibfnamefont {R}~\bibnamefont {Kleiner}},\ }\bibfield  {title} {\enquote
  {\bibinfo {title} {Improving the performance of superconducting microwave
  resonators in magnetic fields},}\ }\href@noop {} {\bibfield  {journal}
  {\bibinfo  {journal} {Applied Physics Letters}\ }\textbf {\bibinfo {volume}
  {98}},\ \bibinfo {pages} {102504} (\bibinfo {year} {2011})}\BibitemShut
  {NoStop}%
\bibitem [{\citenamefont {Graaf}\ \emph {et~al.}(2012)\citenamefont {Graaf},
  \citenamefont {Danilov}, \citenamefont {Adamyan}, \citenamefont {Bauch},\
  and\ \citenamefont {Kubatkin}}]{graaf2012magnetic}%
  \BibitemOpen
  \bibfield  {author} {\bibinfo {author} {\bibfnamefont {SE~de}\ \bibnamefont
  {Graaf}}, \bibinfo {author} {\bibfnamefont {AV}~\bibnamefont {Danilov}},
  \bibinfo {author} {\bibfnamefont {Astghik}\ \bibnamefont {Adamyan}}, \bibinfo
  {author} {\bibfnamefont {Thilo}\ \bibnamefont {Bauch}}, \ and\ \bibinfo
  {author} {\bibfnamefont {SE}~\bibnamefont {Kubatkin}},\ }\bibfield  {title}
  {\enquote {\bibinfo {title} {Magnetic field resilient superconducting fractal
  resonators for coupling to free spins},}\ }\href@noop {} {\bibfield
  {journal} {\bibinfo  {journal} {Journal of Applied Physics}\ }\textbf
  {\bibinfo {volume} {112}},\ \bibinfo {pages} {123905} (\bibinfo {year}
  {2012})}\BibitemShut {NoStop}%
\bibitem [{\citenamefont {Samkharadze}\ \emph {et~al.}(2016)\citenamefont
  {Samkharadze}, \citenamefont {Bruno}, \citenamefont {Scarlino}, \citenamefont
  {Zheng}, \citenamefont {DiVincenzo}, \citenamefont {DiCarlo},\ and\
  \citenamefont {Vandersypen}}]{samkharadze2016high}%
  \BibitemOpen
  \bibfield  {author} {\bibinfo {author} {\bibfnamefont {N}~\bibnamefont
  {Samkharadze}}, \bibinfo {author} {\bibfnamefont {A}~\bibnamefont {Bruno}},
  \bibinfo {author} {\bibfnamefont {P}~\bibnamefont {Scarlino}}, \bibinfo
  {author} {\bibfnamefont {G}~\bibnamefont {Zheng}}, \bibinfo {author}
  {\bibfnamefont {DP}~\bibnamefont {DiVincenzo}}, \bibinfo {author}
  {\bibfnamefont {L}~\bibnamefont {DiCarlo}}, \ and\ \bibinfo {author}
  {\bibfnamefont {LMK}\ \bibnamefont {Vandersypen}},\ }\bibfield  {title}
  {\enquote {\bibinfo {title} {High-kinetic-inductance superconducting nanowire
  resonators for circuit qed in a magnetic field},}\ }\href@noop {} {\bibfield
  {journal} {\bibinfo  {journal} {Physical Review Applied}\ }\textbf {\bibinfo
  {volume} {5}},\ \bibinfo {pages} {044004} (\bibinfo {year}
  {2016})}\BibitemShut {NoStop}%
\end{thebibliography}%

\end{document}